\documentclass[manuscript]{copernicus}

\RequirePackage{amsthm,amsmath,amsfonts,amssymb}
\RequirePackage[authoryear]{natbib}
\RequirePackage{graphicx}
\RequirePackage{nicefrac}
\RequirePackage{tikz}
\usepackage{subcaption}

\def\figdir{./fig}
\newcommand\dint{\mathop{}\!\mathrm{d}}
\def\AA{\mathbf{A}}

\def\HH{\mathbf{H}}
\def\RR{\mathbb{R}}

\def\WW{\mathbf{W}}

\def\ss{\boldsymbol{s}}
\def\xx{\boldsymbol{x}}
\def\yy{\boldsymbol{y}}
\def\lon{\mathrm{lon}}
\def\lat{\mathrm{lat}}
\def\Lon{x_\mathrm{lon}}
\def\Lat{x_\mathrm{lat}}
\def\TextBefore{baseline}
\def\TextAfter{signal}
\def\TempProf{\mathrm{T}}
\def\TempProf{T}
\def\SATempProf{\tilde \TempProf}
\def\TFloat{t_\text{\TextAfter}}
\def\TFloatBefore{t_\text{\TextBefore}}
\def\OrigCoords{((\Lon, \Lat), \TFloat; \TFloatBefore)}
\def\XTrack{d}
\def\DT{\tau}
\def\HurRefCoords{(\XTrack, \DT)}
\def\Pres{z}
\def\YD{t}

\def\XProj{\pi(\Lon, \Lat)}
\def\TProj{t_{\XProj}}
\def\ProjCoords{(\XProj, \TProj)}
\def\TPSFixedKnotBasis{\tilde{\mathbf{E}}}
\def\SeasonalWindowSize{8}
\def\GPWindowSize{5}
\def\TempDiff{y}
\def\TempDiffJ{y^{(ij)}}
\def\TempDiffK{y^{(ik)}}
\def\TempDiffJK{\yy^{(ijk)}}
\def\TempDiffAll{\yy}
\def\TempIJK{\tilde{\boldsymbol{T}}^{(ijk)}}

\def\TempAfterK{\tilde y^{(k)}}
\def\RawTempBefore{\TempProf_\text{\TextBefore{}}}
\def\RawTempAfter{\TempProf_\text{\TextAfter{}}}
\def\TempBefore{\tilde \TempProf_\text{\TextBefore{}}}
\def\TempAfter{\tilde \TempProf_\text{\TextAfter{}}}
\def\TempBeforeI{\tilde \TempProf_\text{\TextBefore{}}^{(i)}}
\def\TempAfterJ{\tilde \TempProf_\text{\TextAfter{}}^{(j)}}
\def\TempAfterK{\tilde \TempProf_\text{\TextAfter{}}^{(k)}}
\def\YDBefore{t_\text{\TextBefore{}}}
\def\YDAfter{t_\text{\TextAfter{}}}
\def\TimeBefore{\YDBefore}
\def\TimeAfter{\YDAfter}
\def\YDBeforeI{t_i}
\def\YDAfterJ{t_j}
\def\YDAfterK{t_k}

\def\GPPhi{\phi}
\def\GPThetaLat{\theta_\lat}
\def\GPThetaLon{\theta_\lon}
\def\GPThetaT{\theta_t}
\def\GPSigma{\sigma}

\DeclareMathOperator{\var}{\mathrm{var}}
\DeclareMathOperator{\cov}{\mathrm{cov}}
\newcommand{\norm}[1]{\lVert#1\rVert}
\newcommand{\bmat}[1]{\begin{bmatrix}#1\end{bmatrix}}

\newcommand{\iid}{\stackrel{\mathrm{iid}}{\sim}}
\DeclareMathOperator{\GP}{\mathrm{GP}}
\def\GPParams{\theta}

\def\tx{\boldsymbol{\xi}}
\def\txi{\xi}
\def\tX{\boldsymbol{\Xi}}
\def\tS{\ss}
\def\tY{\yy}
\def\ty{y}
\def\ts{s}
\def\TPSBasisMatrix{\mathbf{R}}
\def\TPSFixedKnotTestBasis{\hat{\TPSBasisMatrix}}
\def\TPSFixedKnotBasis{\tilde{\TPSBasisMatrix}}
\def\LinBasisMatrix{\mathbf{L}}
\def\LinFixedKnotTestBasis{\hat{\LinBasisMatrix}}

\def\TPSKnotCoefs{\hat{\boldsymbol{\delta}}}

\def\TPSLinearCoefs{\hat{\boldsymbol{\beta}}}
\def\TPSNFKnotCoefs{\boldsymbol{\delta}}
\def\TPSNFLinearCoefs{\boldsymbol{\beta}}

\def\TPSCov{{\boldsymbol{\Sigma}}}

\def\TPSFixedKnotSelfBasis{{\mathbf{S}}}

\def\TCSignal{s(\XTrack, \DT)}
\def\MFBefore{m(\Lon, \Lat, \YD_\text{\TextBefore{}})}
\def\MFAfter{m(\Lon, \Lat, \YD_\text{\TextAfter{}})}
\def\VarBefore{a(\Lon, \Lat, \YD_\text{\TextBefore{}})}
\def\VarAfter{a(\Lon, \Lat, \YD_\text{\TextAfter{}})}
\def\VarBeforeSmall{a(\xx, \YD_\text{\TextBefore{}})}
\def\VarAfterSmall{a(\xx, \YD_\text{\TextAfter{}})}

\def\CVLamb{\hat\lambda^\text{cv}}

\begin{document}

\title{Spatio-temporal methods for estimating subsurface ocean thermal %
response to tropical cyclones}


\Author[1,4]{Addison J.}{Hu}
\Author[1]{Mikael}{Kuusela}
\Author[1,4]{Ann B.}{Lee}
\Author[2]{Donata}{Giglio}
\Author[3]{Kimberly M.}{Wood}

\affil[1]{Department of Statistics \& Data Science, Carnegie Mellon University, USA}
\affil[2]{Department of Atmospheric and Oceanic Sciences, University of Colorado Boulder, USA}
\affil[3]{Department of Hydrology and Atmospheric Sciences, University of Arizona, USA}
\affil[4]{Machine Learning Department, Carnegie Mellon University, USA}




\correspondence{Addison J. Hu (\url{mail@huisaddison.com})}

\runningtitle{Spatio-temporal methods for TC-ocean response estimation}

\runningauthor{A. J. Hu et al.}

\received{}
\pubdiscuss{} 
\revised{}
\accepted{}
\published{}


\firstpage{1}

\maketitle

\begin{abstract}
  Tropical cyclones (TCs), driven by heat exchange between the air and sea,
pose a substantial risk to many communities around the world.  Accurate
characterization of the subsurface ocean thermal response to TC passage
is crucial for accurate TC intensity forecasts and for an understanding of 
the role that TCs play in the global climate system.  
However, that characterization
is complicated by the high-noise ocean environment, correlations inherent
in spatio-temporal data, relative scarcity of \emph{in situ} observations,
and the entanglement of the TC-induced signal with seasonal signals.
We present a
general methodological framework that addresses these difficulties, integrating
existing techniques in seasonal mean field estimation, Gaussian process
modeling, and nonparametric regression into an ANOVA decomposition model.
Importantly, we improve upon past work by properly handling seasonality,
providing rigorous uncertainty quantification, and treating time as a
continuous variable, rather than producing estimates that are binned in time.
This ANOVA model is estimated using \emph{in situ} subsurface
temperature profiles from the Argo fleet of autonomous floats through a
multi-step procedure, which (1) characterizes the upper ocean seasonal shift
during the TC season; (2) models the variability in the temperature
observations; (3) fits a thin-plate spline using the variability estimates to
account for heteroskedasticity and correlation between the observations.  This
spline fit reveals the ocean thermal response to TC passage.  Through this
framework, we obtain new scientific insights into the interaction between TCs
and the ocean on a global scale, including a three-dimensional characterization
of the near-surface and subsurface cooling along the TC storm track and the
mixing-induced subsurface warming on the track's right side.

\end{abstract}


\introduction{
  \label{sect:introduction}
  Tropical cyclones (TCs) occur in most oceanic basins, posing a substantial
risk to many communities across the globe.  This risk is exacerbated in a
warming climate, in which heightened ocean heat content and sea surface
temperatures (SSTs) contribute to stronger and longer lasting TCs, with
intensified flooding from increased atmospheric moisture
\citep{trenberth_hurricane_2018,potter_tropical_2019,
russell_investigating_2020,trepanier_north_2020,xu_tropical_2020}.
Predominant theory holds that TCs are chiefly driven by air-sea
interchange of thermal energy (e.g.,
\citealp{emanuel_air-sea_1986,emanuel_thermodynamic_1999}),
with SSTs above 26.5$^\circ$C conducive to TC intensification
\citep{mctaggart-cowan_revisiting_2015}.  The intensification of TCs
generates several observable physical phenomena, including extraction
of energy from the ocean surface layer and upward mixing of cooler subsurface
water, both of which register as ``cold wakes'', i.e., lower temperatures
compared to the pre-TC passage baseline
\citep{emanuel_contribution_2001,haney_hurricane_2012,
haakman_statistical_2019}.  This
cooling of surface
waters yields a decrease in SST, possibly inhibiting further TC
intensification or even maintenance of current intensity
\citep{bender_real-case_2000,shay_air-sea_2010}.  The subtle interplay
between TC intensification and upper-ocean thermal energy
is therefore important for accurate TC intensity forecasts
\citep{mainelli_application_2008}.

Even after decades of research on TCs, a full understanding of
the physics of the air–sea interface at high wind speeds is missing, which
hinders our ability to simulate the structure and intensity of
hurricane-strength storms \citep{emanuel_100_2018}.
The principal aim of this paper is to provide statistical methodology to
characterize temperature changes in the near-surface and subsurface ocean
associated with the passage of a TC, given temperature observations
from autonomous robotic devices called Argo floats
\citep{woods_hole_oceanographic_institution_argo_2017}.
In a naive sense, one may observe the TC-induced temperature
change merely by comparing Argo temperature values, as a function of
depth, recorded before
and after the passage of a TC, as in Figure~\ref{fig:argo-temp-profile}.
However, the scarcity of such observations makes it nontrivial to obtain
a more detailed characterization of the TC-induced signal.  Here
we develop methodology that enables such a characterization as a function
of perpendicular distance\footnote{
  Measured in terms of angular distance to the nearest point on the TC track;
  alternately referred to as ``cross-track distance'' and ``cross-track
  angle''.
}from the TC track, time since TC passage, and
subsurface depth.  We call this the \emph{TC-induced signal} (or ``TC signal'')
in a \emph{TC-centric coordinate system}.

Challenges that remain in improving estimates of TC-induced ocean temperature
changes include: (1) the high-noise environment of the ocean, coupled with the
relative scarcity of \emph{in situ} observations, which makes it difficult to
observe statistically meaningful effects;  (2)
the need to disentangle the TC-induced signal from other, unrelated signals also
present in the data, such as a seasonal warming effect that occurs in
the ocean during the summer months; and (3) the highly correlated nature of the
spatio-temporal Argo profiles, which necessitates careful estimation of the
covariance between profiles located close to each other in space and time.  In
addressing each of these obstacles posed by the complex data model within which
we work, we obtain an estimate of the TC-induced signal with a reasonable
signal-to-noise ratio.

The underlying statistical framework can be framed in terms of a classical
ANOVA decomposition of each temperature profile into a TC-induced signal,
seasonal mean effect, and ocean variability. At first glance, this appears to
be a spatio-temporal ANOVA model (e.g., \citealp{luo1998spatial}.  However, one
must carefully
note that while the seasonal mean effect and ocean variability are measured in
the geographic coordinates of longitude, latitude, and time of year, the
TC-induced signal is parameterized in the TC-centric coordinate system. These
separate coordinate systems, and the ANOVA decomposition bridging these two
domains, will be formalized in the methodological development. This novel twist
on the usual spatio-temporal ANOVA model allows us to pool TC-related ocean
temperature observations from all TC regions for the entire TC season, and
indeed across TC seasons, while still producing our estimate of the TC-induced
signal in its natural TC-centric coordinate system.

The past work closest to our own is \citet{cheng_global_2015}, which
uses an Argo profile pairing procedure to perform a temporally binned
analysis of thermal changes in depth and cross-track distance.  We
build upon this pairing process, which takes advantage of the highly
correlated nature of spatio-temporal data, to make the following
contributions.  First, we provide methodology
to estimate the seasonal shift in upper ocean temperature
over the timespan of the profile pairs considered, based upon
prior oceanographic work \citep{ridgway_ocean_2002,roemmich_20042008_2009}.  
We find a nontrivial seasonal
effect over the timespan of the profile pairs in our analysis, as most TCs
occur during months when the subsurface water column experiences warming.
Armed with this information, we adjust the temperature differences by
removing the estimated seasonal shift.
The previous analysis of \citet{cheng_global_2015} does not perform this
seasonal adjustment, which may impact their estimate of the TC-related
signal at larger time lags, as the seasonal warming would not be distinguished
from their TC-related signal.

For these seasonally adjusted temperature
differences, we further detail a method for estimating their self-variances and 
cross-covariances, based upon the locally stationary Gaussian process model of
\citet{kuusela_locally_2018}.  The reasons for fitting these covariances prior
to producing the final fits are twofold: ocean dynamics conspire to produce
additional heteroskedastic noise at large time lags; and pairs of profiles,
especially those nearby in space and time, can be highly correlated.
These covariance estimates are then
used in a thin-plate spline smoother to account for the effect of the
correlated, heteroskedastic observations and to provide pointwise confidence
intervals.  Through these steps, we attain the chief
scientific contributions of characterizing the ocean response to TC
passage in the continuous time realm, rather than using a binned analysis
as in
\cite{cheng_global_2015}, as well as properly accounting for the seasonal
variation in ocean temperatures over the time scales of the paired profiles.
Moreover, our estimation of covariances between profiles co-located in space
and time,
which were not accounted for in \citet{cheng_global_2015}, admits more
accurate
fits and enables rigorous uncertainty quantification.  These
statistical procedures are unified in terms of an ANOVA decomposition,
which is fit separately over a grid of twenty depth levels.

Through this model, we more accurately characterize ocean cooling
in the near-surface and subsurface waters, as well as the mixing induced warming
in the subsurface on one side of the TC center.
For a fixed depth, this characterization of the TC signal is performed by
estimating a
continuous function of cross-track distance and time since TC passage.  We then
repeat this estimation over a fine grid of depths from the ocean
surface to 200 meters below the surface, using Argo profiles and TC track
data from 2007 to
2018.  In particular, since our estimates of the temperature change
vary continuously over these
three axes (cross-track distance, time since TC passage, and depth), we are
able to produce a three-dimensional characterization of the aforementioned
scientific phenomena.

The organization of the paper is as follows.
Section~\ref{sec:scientific_context} formally motivates,
in oceanographic, meteorological, and climatological terms,
the scientific
problem of estimating the ocean thermal response to tropical cyclone
passage.
Section~\ref{sect:data}
introduces the tropical cyclone track data and the Argo profile data.
Section~\ref{sect:methods} describes our methodology, where we begin by
presenting the ANOVA-type data model and
its implied data analysis pipeline.  Data
preprocessing
is reviewed in Section~\ref{sub:argo_profile_preprocessing}, and
the process for pairing
profiles
and projecting pairs onto TC tracks is detailed in
Section~\ref{sect:pairing-process}.  A local mean field model for
capturing TC-unrelated seasonal effects is described in 
Section~\ref{sect:mean-field}, followed by a Gaussian process model
for estimating ocean variability
in Section~\ref{sect:gaussian-process}.  In
Section~\ref{sect:thin-plate-splines}, we describe a version of the thin-plate
spline that allows for covariance reweighting.
Section~\ref{sect:application}
describes the results of applying our methodology to the estimation of the
global ocean
thermal response to TCs.  We conclude with a discussion of the results and
future directions in Section~\ref{sect:discussion}.  

Appendix~\ref{app:thinplatespline} provides a comprehensive review
of the motivation and derivation for the thin-plate spline variant used,
and Appendix~\ref{app-subsect:leave_one_out_cross_validation} details the
leave-one-out cross-validation procedure used to select the regularization
parameter.  
A full set of model fits is provided in the online supplement
\citep{spatiotemporal_tc_methods_supplement}.
In the spirit of reproducibility and to encourage the extension of these
results,
the code used to process the data and fit these models is freely available
online at \url{https://github.com/huisaddison/tc-ocean-methods}.

}

\section{Scientific context}%
\label{sec:scientific_context}
A prevailing theory states that tropical cyclones are primarily driven by
two physical phenomena: (1) energy flux from the ocean and (2) the
temperature difference between the ocean surface and the tropopause, the
top boundary of the lowest layer of Earth's atmosphere,
in a
process called wind-induced surface heat exchange (WISHE;
e.g., \citealp{emanuel_air-sea_1986, emanuel_thermodynamic_1999}).  SSTs of at
least
$26.5^\circ$C are generally sufficient to support TC genesis
\citep{mctaggart-cowan_revisiting_2015}, and additional energy becomes
available to the TC as SSTs increase beyond this threshold.  

As TCs intensify, their strengthening surface winds
extract energy from the ocean surface and induce mixing of the upper
ocean.  This combination of processes decreases
SST (creating a cold wake) and thus decreases the energy
available for further TC intensification 
\citep{bender_real-case_2000,shay_air-sea_2010}.
The magnitude of mixing depends on the strength of the
TC's surface winds, its size, and translation speed, as well as background
ocean conditions.
Therefore, knowledge
of subsurface ocean conditions can
impact the accuracy of TC
intensity forecasts (e.g., \citealp{mainelli_application_2008}).
Such information, however, is difficult to obtain through remotely sensed,
satellite-based
observations, necessitating the introduction of \emph{in situ}
measurements, such as those provided by the Argo float program
\citep{woods_hole_oceanographic_institution_argo_2017}.

Latent heat, the energy needed to evaporate liquid water and released
when water vapor condenses, is the primary mechanism by which energy
is transferred from the ocean to the atmosphere.  Through latent heat
transfer, the passage of a TC induces
temperature decreases at the ocean surface and within the
ocean mixed layer
\citep{fisher_exchange_1957,leipper_observed_1966,bender_real-case_2000,
dasaro_cold_2007,lin_interaction_2005,lloyd_observational_2011,
balaguru_ocean_2012,elsberry_mixed_1976,price_upper_1980}.
Observational studies have examined the thermodynamic response
of the ocean surface to TC passage via observations from
survey ships or expendable bathythermographs
\citep{price_upper_1980,shay_effects_2000,cione_sea_2003,
dasaro_cold_2007,dare_sea_2011,haakman_statistical_2019}, especially before
the Argo array of profiling floats was deployed.  

With more than two million profiles since the early 2000s, Argo floats provide
unprecedented spatial and temporal coverage of the global ocean in the upper
2000 meters
\citep{roemmich_20042008_2009,riser_fifteen_2016,
woods_hole_oceanographic_institution_argo_2017}.  Due to their autonomous
nature, Argo floats are well-positioned to sample the ocean state in three dimensions
before and after TC passage.
Several previous studies have leveraged Argo
profiles to describe TC-related changes in the upper ocean
\citep{liu_upper_2007,balaguru_dynamic_2015,
sun_ocean_2012,qu_sea_2014,cheng_global_2015,lin_ocean_2017,
steffen_barrier_2018,trenberth_hurricane_2018}, focusing on changes in ocean
temperature, salinity, or both, in order to analyze changes in ocean density.
The closest work to ours is \cite{cheng_global_2015}, who perform a binned
analysis of temperature changes in time
(compared to our continuous treatment of time) and do not
account for a seasonal warming effect we discern in our analysis.  An earlier
paired analysis in the north Pacific was performed by \citet{park2011argo}.

Processes regulating the thermal recovery of cold ocean wakes associated with
the passage of a TC depend on the parameters of the wake, e.g., depth, width,
and wind stress \citep{haney_hurricane_2012}. 
Once a TC passes, thermal recovery
results in a net warming of the water
column, a warming which increases with TC intensity \citep{mei_sea_2013}.
Globally,
this net ocean warming is on the order of 0.5 PW
\citep{sriver_investigating_2008,mei_restratification_2012,
mei_spatial_2013,mei_sea_2013},
equivalent to approximately $\nicefrac{1}{6}$ of the ocean heat transport out of the
tropics \citep{wunsch_past_nodate}.  In the long-term mean, this added heat
is either advected poleward by ocean currents
\citep{emanuel_contribution_2001,korty_tropical_2008} or eventually released
into the atmosphere during subsequent cold seasons
\citep{pasquero_tropical_2008,jansen_seasonal_2010}.  For stronger TCs,
part of the extra ocean warming occurs below the winter mixed-layer depth
\citep{mei_sea_2013} and can influence global meridional heat transport.
However,
the transport of this heat may be primarily equatorward
\citep{jansen_impact_2009}, in which case the heat transport would
not contribute to the redistribution of heat within the climate
system from the warmer tropics to the colder poles.
The methodological framework presented in this
manuscript provides an opportunity to contribute to this ongoing discussion
because it helps characterize the evolution of TC-related ocean
temperature changes in greater detail, accuracy, and rigor than previously
available.

\section{Data}\label{sect:data}
The methodology described in this paper requires two main sources of data:
TC track records and subsurface temperature profile databases, both
described in the following subsections.

\subsection{Tropical cyclone track data}
We obtain TC \emph{best-track} data from the National Hurricane Center
(NHC) and the United States Navy Joint
Typhoon Warning Center (JTWC).
Best-track data are produced in post-season analysis using all available
observations, e.g., satellite microwave imagery and aircraft
reconnaissance, to generate the best estimate of
TC location and intensity. Data
are six-hourly with occasional extra time points for landfall.
The NHC Hurricane Database (HURDAT2) provides
best-track data for the North Atlantic and Eastern Pacific basins
\citep{landsea_atlantic_2013}.  The best-track data for the Western Pacific
basin, North Indian Ocean basin, and basins
in the Southern Hemisphere
are obtained from the JTWC \citep{chu_joint_2002}.
Together, these five regions capture all TC activity on Earth.  We
chose to use best-track data from the NHC and the JTWC because both
agencies provide TC intensity estimates based on a common
definition of 1-minute,
10-m maximum sustained wind, rounded to the nearest five knots.  A final,
important
remark when considering data from the different ocean basins is the fact
that aircraft reconnaissance is present in the North Atlantic but largely
absent in the other basins, likely providing more accurate TC intensity
estimates in the North Atlantic compared to elsewhere.

We consider all TCs from 2007 to 2018.  The start year of 2007 is chosen
to ensure sufficient coverage by Argo floats, detailed in the following
section.  Over this span of 12 years, we have 1089 total tracks, with {191} in
the North Atlantic; {223} in the Western Pacific; {339} in the Eastern Pacific;
{66} in the North Indian Ocean; and {270} in the Southern Hemisphere.  These
tracks
are depicted in Figure~\ref{fig:global-tracks}.

\subsection{Argo float data}
\begin{figure}
	\begin{subfigure}{.53\textwidth}
		\centering
    \vspace{0.30cm}
		\includegraphics[width=1.\linewidth]{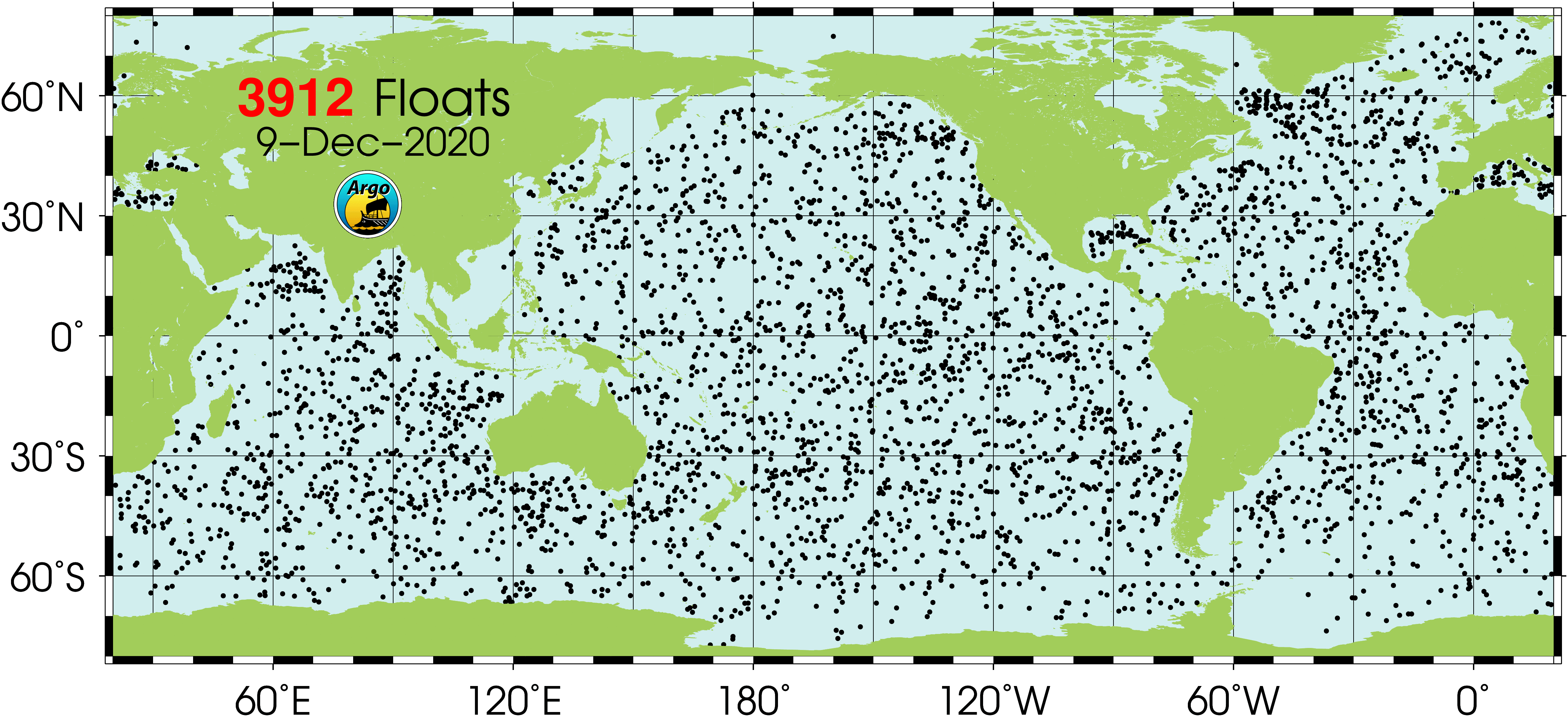}  
    \vspace{0.15cm}
    \caption{Global coverage of Argo floats
      \citep{argo_program_status}.
          }
		\label{fig:argo-profiles-global}
	\end{subfigure}
  \hspace{0.3cm}
	\begin{subfigure}{.40\textwidth}
		\centering
		\includegraphics[width=1.\linewidth]{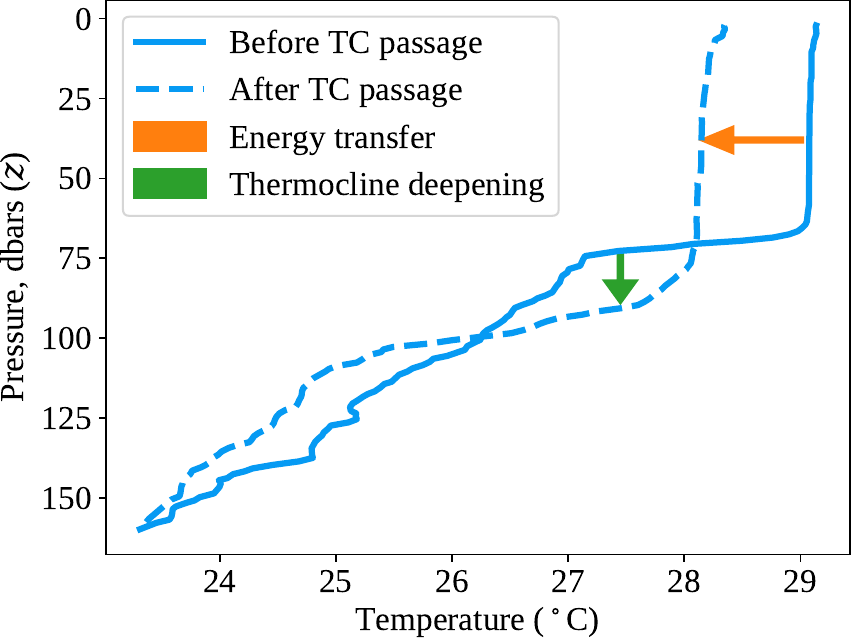}  
		\caption{Argo temperature profiles}
		\label{fig:argo-temp-profile}
	\end{subfigure}
    \caption{
        Argo floats have achieved global coverage since 2007
        (Figure~\ref{fig:argo-profiles-global}).
        Figure~\ref{fig:argo-temp-profile}
        shows two consecutive temperature profiles from the same Argo
        float near Puerto Rico in 2017: one before and one after the passage
        of
        Hurricane Maria.  The orange arrow depicts the net removal of
        thermal energy from the ocean, while the green arrow illustrates
        the deepening of the thermocline as a result of increased mixing.
    }
	\label{fig:argo}
\end{figure}

Argo is a network of autonomous floats sampling global subsurface ocean
temperature and salinity, with each float reporting roughly every ten days.
Argo floats have been deployed since the early 2000s and reached the desired
nominal distribution of one float for every $3^\circ$ longitude by $3^\circ$
latitude box in the global ice-free ocean in 2007
\citep{woods_hole_oceanographic_institution_argo_2017},
providing unprecedented subsurface spatial and temporal coverage,
with no seasonal sampling bias \citep{roemmich_20042008_2009}.   The
geographical
distribution of Argo floats is depicted in
Figure~\ref{fig:argo-profiles-global}.
Each Argo observation consists of temperature and salinity measurements,
indexed by
pressure, taken as the Argo float ascends to the surface
from a depth of roughly 2000 meters (m).
The resulting vertical profile is associated with a timestamp as well as
satellite-determined latitude and longitude coordinates.  
Figure~\ref{fig:argo-temp-profile} plots two such temperature profiles,
reported in roughly the same location before and after the passage of
Hurricane Maria (2017).  The leftward shift from the first profile to the
second (denoted with an orange arrow) illustrates the reduction in thermal
energy associated with Hurricane Maria's passage.  The green arrow
illustrates the deepening of the \emph{thermocline}, the change point between
an upper, well-mixed layer and cooler subsurface waters, as a result
of increased mixing.
For the rest of the paper, we will interchangeably use the terms ``depth,''
which is more familiar, and ``pressure,'' which is more accurate. Argo floats
record pressure in decibars (dbar), and one decibar
change in pressure corresponds to approximately one meter change in depth
\citep{talley_descriptive_2011}.

For our purposes, the chief contribution of Argo to our oceanographic and
meteorological pursuits is its provision of vertical information, from
the surface of the ocean to roughly 2000 m below the surface,
in tandem with its global coverage, with one float per $3^\circ$ longitude
by $3^\circ$ latitude box on average
\citep{woods_hole_oceanographic_institution_argo_2017}.
The vertical information from the Argo profiles, unattainable by remote
sensing methods such as satellites, is key to quantifying the
thermal response of the ocean to the passage of a TC, which varies with depth.
Argo profiles, however, are not without shortcomings for our purposes; their
spatio-temporal resolution is coarser than typical TC space-time scales,
which necessitates the statistical methods developed herein.

An Argo float takes a profile in the following fashion.  Every sampling period,
typically every 10 days, the float descends from its ``parking'' depth of
1000 m to a profiling depth of 2000 m, and then ascends from 2000 m to the
surface over the course of about six hours, taking observations of
temperature and salinity along the way
\citep{woods_hole_oceanographic_institution_argo_2017}.   
It then rests at the surface for 15 minutes to an hour, determining its
location and communicating its measurements via satellite.  Finally, the float
returns to its parking depth until its next sampling period,
to minimize float drift.

In this study, we refer to an individual Argo
device as a ``float'' and to the measurements taken vertically at a certain
longitude, latitude, and time
as a ``profile.''  Each
Argo profile is uniquely identified by its float identifier (unique to an Argo
float) and cycle number (for a given float, unique to each profile).

\subsection{TC-centric coordinate system}
\begin{figure}
	\begin{subfigure}{.53\textwidth}
		\centering
		\includegraphics[width=1.\linewidth]{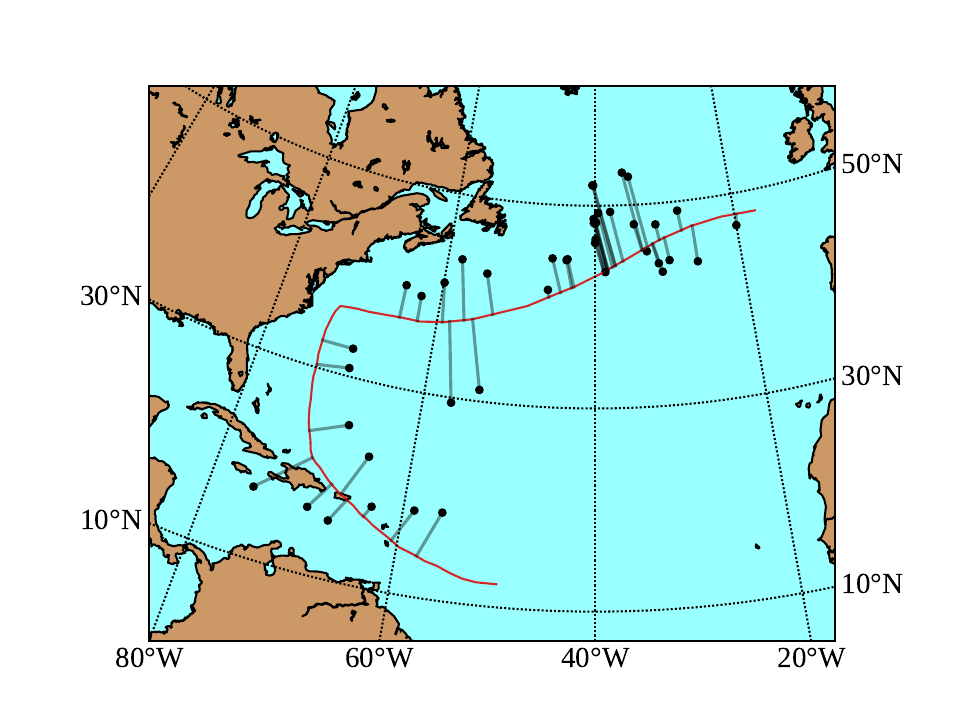}  
		\caption{
            Representation in $(\Lon, \Lat)$ coordinates
        }
		\label{fig:hurricane-maria-lon-lat}
	\end{subfigure}
	\begin{subfigure}{.40\textwidth}
		\centering
		\includegraphics[width=1.\linewidth]{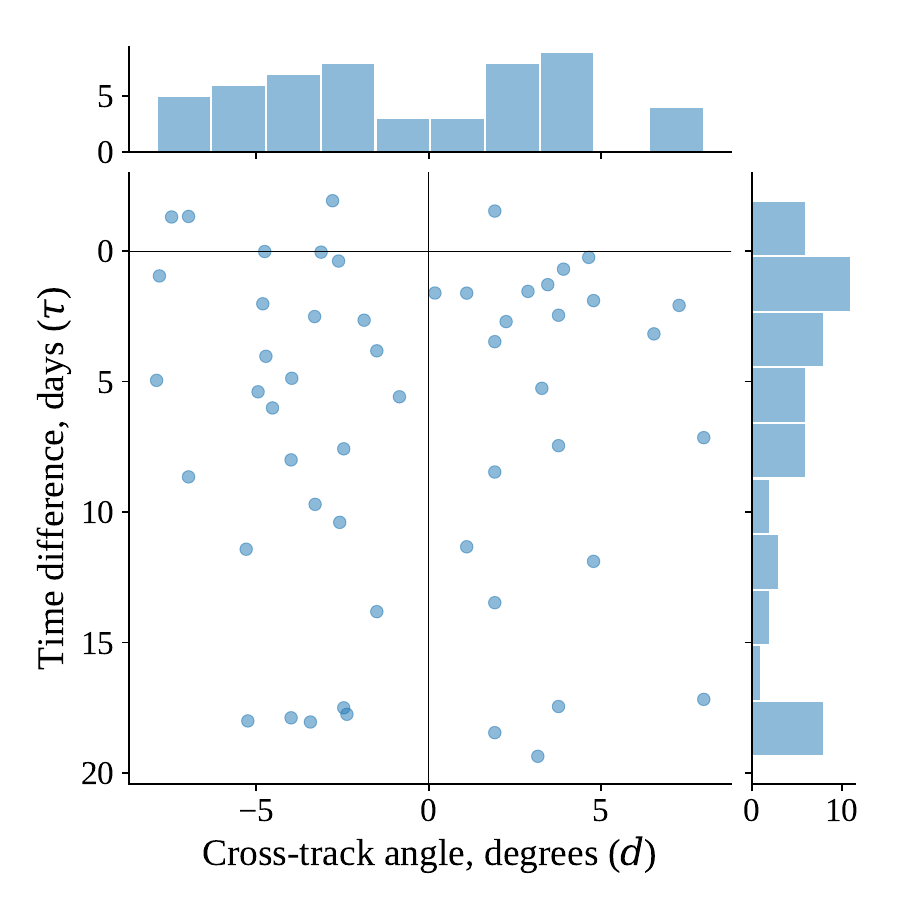}  
		\caption{
            Representation in $(\XTrack, \DT)$ coordinates
        }
		\label{fig:crosstrack-maria-profiles}
	\end{subfigure}
    \caption{
        Illustration of the coordinate systems with Hurricane Maria
        (2017).
        Figure~\ref{fig:hurricane-maria-lon-lat}, using the latitude-longitude
        coordinate system,
        shows the path of the storm
        in red and the location of the Argo profile pairs straddling the
        storm passage in black.
        In Figure~\ref{fig:crosstrack-maria-profiles}, we
        plot those same pairs of profiles in the TC-centric coordinate system.
        It is possible for one black point in the latitude-longitude
        coordinate system to correspond to multiple observation in the
        TC-centric coordinate system, if a single profile occurring before
        TC passage is paired with multiple profiles occurring after TC passage.
    }
	\label{fig:hurricane-maria}
\end{figure}

\begin{figure}
	\begin{subfigure}{.55\textwidth}
    \vspace{1.25cm}
		\centering
		\includegraphics[width=1.\linewidth]{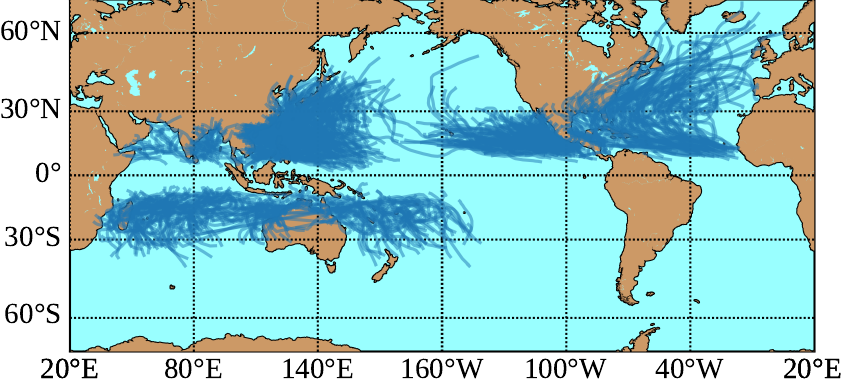}  
    \vspace{0.43cm}
		\caption{
            Representation in $(\Lon, \Lat)$ coordinates,\\
            1089 tracks in total.
        }
		\label{fig:global-tracks}
	\end{subfigure}
  \hspace{0.5cm}
	\begin{subfigure}{.40\textwidth}
		\centering
		\includegraphics[width=1.\linewidth]{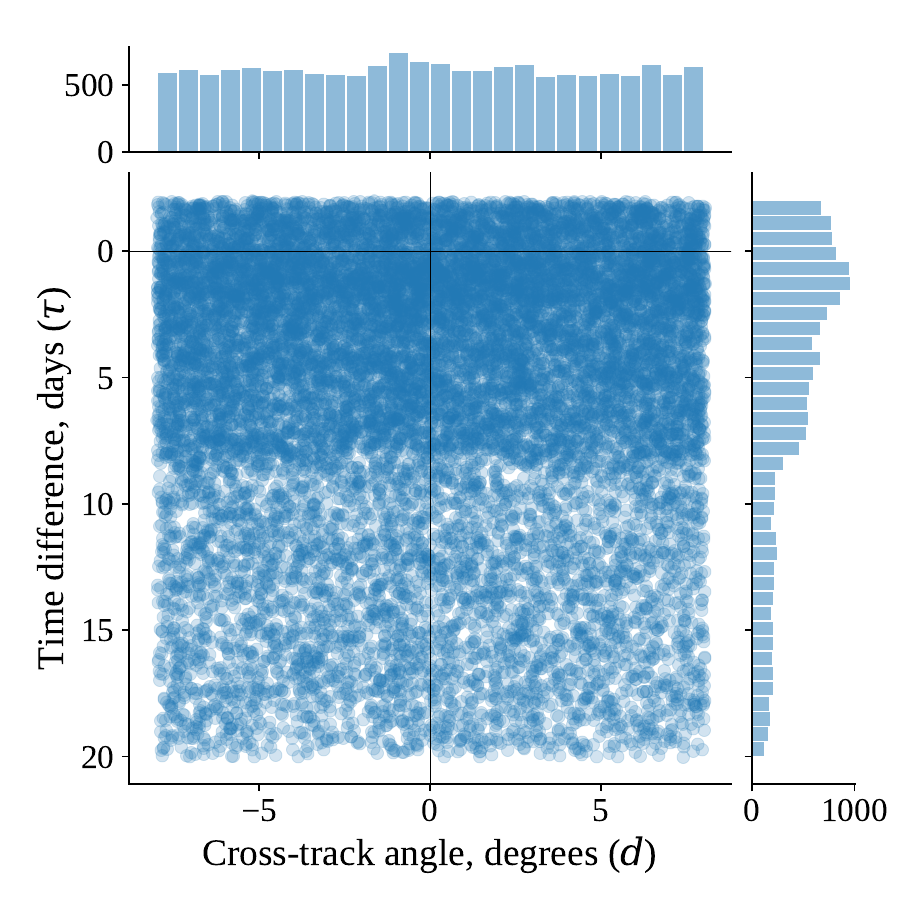}  
		\caption{
      Representation in $(\XTrack, \DT)$ coordinates, 16025 pairs total.
    }
		\label{fig:crosstrack-profiles}
	\end{subfigure}
  \caption{
      Global tropical cyclones from {2007} to {2018} in the latitude-longitude
      coordinate system (left panel) and in the TC-centric coordinate
      system (right panel).
      Marginal histograms illustrate the density of the pairs along
      the time and cross-track axes.
  }
	\label{fig:global-tracks-profiles}
\end{figure}

To quantify the ocean thermal response to the passage of a TC,
we perform a pairing procedure inspired by \cite{cheng_global_2015}.  
Intuitively, we seek pairs of profiles which ``straddle'' the passage of TC,
i.e., one observation before and one observation during or after the TC passage.
These Argo profile pairs are placed into a TC-centric coordinate system through
a projection process, which
rewrites the space-time coordinates of an Argo profile pair in reference to
a passing tropical cyclone.  Figure~\ref{fig:hurricane-maria} illustrates
this process using the TC track and profile pairs associated with
Hurricane Maria (2017), and the full set of profile pairs in the TC-centric
coordinate system is depicted in
Figure~\ref{fig:crosstrack-profiles}.
The pairing and projection procedures are formalized in
Section~\ref{sect:pairing-process}.

\section{Statistical methodology}\label{sect:methods}
The methodology introduced in this paper aims to produce a representation
of the TC-induced change in ocean temperature, at a sequence of fixed
depths, as a continuous function of time since tropical cyclone passage and
cross-track distance from the center of the tropical cyclone path.  To properly
characterize these temperature changes, we combine a variety of techniques for
pairing
Argo profiles, estimating seasonal effects, modeling space-time
covariances, and smoothing in multiple dimensions.

All notation will be formally introduced in their requisite subsections,
but in general $\xx$ refers to a location in the latitude-longitude 
coordinate system;
$\YD$ refers to day of year; $\XTrack$ refers to the cross-track angle
between a
location and the center of the storm path; and $\DT$ refers to the
time difference between a post-storm Argo profile and the passage of the
tropical cyclone with which is it associated, measured in days.

Before the detailed exposition, we provide a high-level explanation of our
modeling approach, which will be further formalized as an ANOVA decomposition 
data model.  First, we partition the Argo profile database into
``TC profiles'', which report timestamps immediately
before and after the passage of a TC, and
``non-TC profiles''
which are bounded away from TCs in space and time
(Section~\ref{sub:tc_profiles_and_non_tc_profiles}).  Using the
TC profiles, we form in Section~\ref{sect:pairing-process} pairs of Argo
profiles, in which the two profiles ``straddle'' the passage of a TC, i.e.,
one profile occurs
before (the ``\TextBefore{}'' profile) and one profile
occurs during or after the TC passage (the
``\TextAfter{}'' profile).  Then, we
decompose the observed temperature difference (at each pressure level) between
the \TextBefore{} and \TextAfter{} profiles in the following
manner:
\begin{equation}\label{eq:high-level-model}
    \text{Temperature difference}
    =
    \text{TC signal}
    + \text{Seasonal effect}
    + \text{Ocean variability}.
\end{equation}
The nonrandom \emph{seasonal effect} term
is estimated in Section~\ref{sect:mean-field} and is subtracted from
the temperature differences to produce seasonally adjusted temperature
differences.  The \emph{ocean variability} (noise) term is then estimated
in Section~\ref{sect:gaussian-process}; it is the only term on the right-hand
side regarded as random.  These two terms describe the state of the ocean
under the assumption that there are no TC-induced effects; therefore,
partitioning of Argo profiles into TC- and non-TC profiles in Section
\ref{sub:tc_profiles_and_non_tc_profiles} is crucial to their estimation,
as the latter observe the ocean in the TC-absent regime.

The key subject of interest is the \emph{TC signal} term, which we denote by
$\TCSignal$ to emphasize its
dependence on
the cross-track distance from the storm center and the time since storm
passage.
Our goal is to make inferences for functionals $H(s)$ of $s$.  This includes
point evaluators, e.g., $H(s) = s(\XTrack_0, \DT_0)$, as
in Section~\ref{sect:app-tps}, and binned quantities, e.g.,
$H(s) = \frac{1}{\DT_1 - \DT_0} \int_{\DT_0}^{\DT_1} s(\XTrack_0, \DT)
\dint\DT$,
as in Section~\ref{sect:app-depth-cross-track}.

To elucidate the estimation of $s(\XTrack, \DT)$, we describe each of the terms in
\eqref{eq:high-level-model} formally, in terms of an ANOVA decomposition.
Let $\TempProf$ denote an Argo temperature observation, evaluated at a fixed
pressure level $\Pres$.  The \emph{temperature difference} is
the difference between two terms
$\TempProf_\text{\TextAfter{}}
  - \TempProf_\text{\TextBefore{}}$:
\begin{align}
    \label{eq:tempbefore-formal-decomp}
    \TempProf_\text{\TextBefore{}}(\Lon, \Lat, \YD_\text{\TextBefore{}})
        &=  \MFBefore 
            + \VarBefore \\
    \label{eq:tempafter-formal-decomp}
    \TempProf_\text{\TextAfter{}}(\Lon, \Lat, \YD_\text{\TextAfter{}};
          \XTrack, \DT)
        &=  \TCSignal  +  \MFAfter
        + \VarAfter 
\end{align}
where $m$ is a seasonal mean field and $a$ is a Gaussian process term
describing non-TC ocean variability, both
elucidated fully in Sections~\ref{sect:mean-field} and
\ref{sect:gaussian-process}.

A few remarks are in order.  First, notice that $\TempProf_\text{\TextAfter{}}$
is
written as a function both of Earth coordinates $(\Lon, \Lat, \YD)$ as well as
in terms of TC-centric coordinates $(\XTrack, \DT)$, which are uniquely
determined
by the Earth coordinates and the associated TC path.  This is no accident;
while $m$
and $a$ are estimated and only have meaning in terms of the Earth coordinates,
$s$ is estimated and only has meaning in terms of the TC-centric coordinates.
Second, we observe that in theory, \eqref{eq:tempafter-formal-decomp}
suggests that one may directly estimate $s(\XTrack, \DT)$ using only
observations of $\TempProf_\text{\TextAfter{}}$ (and the non-TC profile set).
Unfortunately, one finds that the magnitude
of $a$ is such that estimating $s$ solely using $\TempProf_\text{\TextAfter{}}$ is difficult,
due to a low signal-to-noise ratio.  By taking the temperature difference
$\TempProf_\text{\TextAfter{}} - \TempProf_\text{\TextBefore{}}$, we take advantage of the high correlation
between $\VarAfter$ and $\VarBefore$
to yield results with considerably better signal-to-noise ratio.

\begin{figure}[ht!]
    \centering
    \includegraphics[width=1.0\linewidth]{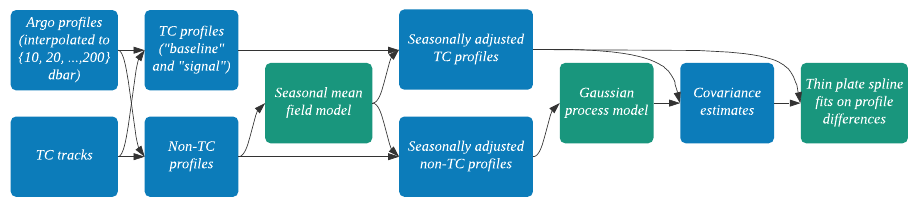}
    \caption{
				Illustration of the multi-step framework as a data analysis pipeline.
        The data and results are denoted in blue, and the statistical models
        are denoted
        in green.  
		}
    \label{fig:pipeline-chart}
\end{figure}

The partitioning of the Argo profile database into TC and non-TC profile
sets
is described in Section~\ref{sub:tc_profiles_and_non_tc_profiles}.
The pairing process and 
the construction of the observed temperature difference are detailed in
Section~\ref{sect:pairing-process}.  Section~\ref{sect:mean-field} explains the
estimation of the seasonal effect based on spatio-temporal local regression
modeling, and Section~\ref{sect:gaussian-process} describes a space-time
Gaussian process model for the remaining
ocean variability term.  Finally, Section~\ref{sect:thin-plate-splines}
leverages thin-plate splines to produce a final fit of the TC signal
$s(\XTrack, \DT)$.
We depict the full framework as a data analysis pipeline in
Figure~\ref{fig:pipeline-chart}.

\subsection{Argo profile preprocessing}
\label{sub:argo_profile_preprocessing}

In this section, we describe two key preprocessing steps performed on
the Argo profiles.

\subsubsection{Interpolation and averaging}
\label{ssub:pchip_interpolation}

An unprocessed Argo temperature profile consists of a sequence of
$\{(\Pres_i, T_i)\}$, which relate pressure level $\Pres_i$ to temperature
measurements $T_i$.  We take each Argo temperature profile and
interpolate it from a pressure level of $\Pres=10$
dbar to $\Pres=200$ dbar using a piecewise cubic hermite interpolating
polynomial (pchip interpolant; \citealp{fritsch_monotone_1980}).
We then read off twenty equally spaced values from $\Pres=10$ dbar 
to $\Pres=200$ dbar,
including the endpoints. We also compute the average temperature
by performing a trapezoidal integration of the pchip over a grid of 5000
evaluations, normalized by the length of integration (190 dbar).  We refer
the former as the \emph{gridded} temperature values and to the latter as the
\emph{vertically averaged} temperature values.

In the analysis that follows, we use the term \emph{Argo profile} to refer
to the collection of the twenty gridded temperature values and the vertically
averaged temperature value.  The partitioning, pairing, and projection
procedures of Sections~\ref{sub:tc_profiles_and_non_tc_profiles} and
\ref{sect:pairing-process} operate on these collections of values on
the profile level.  The statistical methodology, developed in
Sections~\ref{sect:mean-field}, \ref{sect:gaussian-process}, and
\ref{sect:thin-plate-splines}, considers each of the gridded pressure
levels (and the vertically averaged temperature values) separately,
i.e., the statistical analysis is performed twenty-one times, for each
of $z = 10, \dotsc, 200$ and for the averaged temperatures.

We use
$\TempProf$ to denote a single temperature value, either the
evaluation of the pchip interpolant at a pressure level $\Pres$ or the
vertically averaged value.  As the methodology is identical for each
pressure level and for the vertically averaged profiles, we do not burden
$\TempProf$ with additional notation; which of the gridded and vertically
averaged temperature values it is will be made clear from the context.

\subsubsection{TC profiles and non-TC profiles}
\label{sub:tc_profiles_and_non_tc_profiles}

We further partition the Argo profile database into two sets, which
separately characterize the ocean state in the presence and absence of
TCs.  If an Argo profile is within $8^\circ$ (latitude and longitude) of a
tropical cyclone
observation location and within $12$ days before to $20$ days
after that tropical cyclone observation, we denote it a
\emph{tropical cyclone profile} (TC profile).  Any profile not in this set
is denoted a \emph{non-tropical cyclone profile} (non-TC profile).
Though the non-TC profiles do not carry any of the TC signal
$s(\XTrack, \DT)$, they play a crucial role in estimating the
seasonal effect $m$ and ocean variability $a$, which characterize
the ocean state and dynamics under the important assumption that there are
no TC-related effects.  These models are fully described in
Sections~\ref{sect:mean-field} and \ref{sect:gaussian-process}.

\subsection{Profile-TC association process}\label{sect:pairing-process}
To associate Argo profiles with tropical cyclones, we perform a two-step
procedure of \emph{pairing} and \emph{projection}.  In the pairing step,
we identify Argo profiles that are close to each other and straddle
a passing tropical cyclone such that one profile is observed before the
storm to characterize the pre-storm ocean state (referred to as the
\emph{\TextBefore{}} profile) and another is observed
during or after the storm to characterize the change in ocean state
(referred to as the \emph{\TextAfter{}} profile).  In the
projection step, we change the coordinate system of the Argo profile pairs
to one parameterized in terms of the passing tropical cyclone.  This is done by
performing an orthogonal projection of the Argo profile pair onto the tropical
cyclone track.  Both procedures are formalized in the sequel.

We first establish notation. Define the \emph{Earth coordinates} to be
$\OrigCoords$, i.e., the longitude, latitude, and time that the
\TextBefore{} and \TextAfter{} profiles were
taken; and the \emph{TC-centric coordinates} to be $\HurRefCoords$, where
$\XTrack$ is the
signed great circle angular distance on Earth between $(\Lon, \Lat)$ and
$\XProj$, the orthogonal
projection onto the tropical cyclone track, and $\DT$ is the time difference
between $\TFloat$ and $\TProj$, the time when the tropical cyclone was
at $\pi(\Lon, \Lat)$.  Due to the closeness constraints that we impose 
between the \TextBefore{} and \TextAfter{} profiles, described in the following
subsection, only one set of longitude-latitude coordinates is necessary for
both profiles.

\subsubsection{Pairing process}
We follow closely the pairing process introduced in \cite{cheng_global_2015}.
Recall that the tropical cyclone tracks are available at six-hourly time 
resolution and can be made continuous in time and space through straight line
interpolation.  Along each of
these TC tracks, we seek Argo profiles that reported within
$\pm 8^\circ$ (longitude/latitude) and within $12$ days before to $20$
days after each location
on the tropical cyclone track.  We partition these profiles into the
\emph{\TextBefore{}} set ($-12$ to $-2$ days relative to the TC) and the
\emph{\TextAfter{}}
set ($-2$ to $+20$ days relative to the TC).  These two sets contain the
profiles that
are incidental to the TC.  The \TextBefore{} and \TextAfter{} time ranges are
chosen primarily to accommodate sampling constraints posed by the Argo
network; specifically, most floats sample on a ten-day frequency.  
Therefore, the \TextBefore{} time range of $-12$ to $-2$ days ensures that each
float can contribute at least one \TextBefore{} profile.  The \TextAfter{} time
range allows for at least two \TextAfter{} profiles from a given float
(assuming they fit a nearness condition detailed in the sequel).  The
\TextAfter{} profiles begin at $-2$ days to hedge against storm effects prior
to passage and is consistent with the pairing process introduced by
\cite{cheng_global_2015}.

Among profiles incidental to a given TC, we perform a pairing
process as follows.  If there exist a profile in the \TextBefore{} set and a
profile in the \TextAfter{} set that are at most $0.2^\circ$ apart\footnote{
    The $0.2^\circ$ is measured in the induced angle between points on the
    Earth's surface and the
    center of the Earth, and is invariant to the location on Earth at which
    the angle is calculated.  This corresponds roughly to a maximum distance
    of 22 kilometers.  For reference, the JTWC considers TCs of radius 
    $<0.2^\circ$ to be ``very small''
    \citep{JTWC_frequently_nodate}.
}, we pair
them, yielding a \emph{baseline profile} and \emph{signal profile} pair.  Under
this closeness restriction, we make the simplifying assumption that
the \TextBefore{} and \TextAfter{} locations are spatially identical, and for each
pair we use the \TextBefore{} location as the reference point.  We defer a
discussion of how to potentially
extend our model to account for Argo float movement and rigorously incorporate
intra-pair spatial differences to Section~\ref{sect:discussion}.

We impose an additional technical restriction with respect to the
\emph{lineage} of
a \TextBefore{} profile, which is defined to be the set of all \TextAfter{} profiles
associated with a distinct \TextBefore{} profile and that \TextBefore{} profile itself.
We require that all profiles in the lineage of a \TextBefore{} profile must be
separated in time
by at least 72 hours, sparsifying the lineage.
To our knowledge, \cite{cheng_global_2015} imposes no such constraint.
The choice of this separation is for a particular reason:
we know that pairs of profiles close in time tend to be highly correlated, but
our Gaussian process model \citep{kuusela_locally_2018} for estimating the
correlations between pairs of profiles that are close in space and time (a
prerequisite for properly handling them in the final thin-plate spline
fit) may not provide reliable estimates in the rare cases where profiles
occur more frequently than three days apart.
Following a filter on minimum time separation within each lineage, we retain
16025 profile pairs across all five ocean regions: {2757} in the North Atlantic,
{2897} in the East Pacific, {778} in the Indian Ocean, {2185} in the
Southern Hemisphere, and {7409} in the West Pacific.

\subsubsection{Projection process}
Given these Argo profile pairs, which we know straddle a TC path, we require a
procedure for precisely describing the profile
pair's location in space and time relative to the passage of the tropical
cyclone.  Recall that we formally denote the location of the Argo profile
pair by $\OrigCoords$, and that there is no distinction between the
\TextBefore{} and \TextAfter{} locations, as we assume they are the same.
Through our pairing procedure, we rewrite the coordinates of the Argo profile
pair with respect to a nearby tropical cyclone in terms of $\ProjCoords$.

The projection procedure proceeds as follows.  Each tropical cyclone track
is made continuous through straight line interpolation.  We project each
profile pair from $(\Lon, \Lat)$ to $\XProj$, the nearest point on the tropical
cyclone track in the Euclidean sense.  Given $\XProj$, we obtain $\TProj$,
the time at which the tropical cyclone was at $\XProj$, using linear
interpolation.  Finally, we obtain the cross-track angle
\begin{equation}
  \label{eq:xtrack-def}
    \XTrack \triangleq \angle((\Lon, \Lat), \XProj)
\end{equation}
and the time since passage
\begin{equation}
  \label{eq:dt-def}
    \DT \triangleq \TFloat - \TProj
\end{equation}
where $\angle(\cdot, \cdot)$ is the induced angle between 
two points on the Earth's surface with respect to its center, and the
subtraction in \eqref{eq:dt-def}
is performed in a ``year-aware'' fashion, i.e., if the yearday difference
is negative, as may happen over the new year, then we add the number of days
in the prior year to the difference.
The projection process is depicted in Figure~\ref{fig:hurricane-maria}, using
profile pairs associated with Hurricane Maria (2017) as an example.

\subsection{Seasonal mean field estimation}\label{sect:mean-field}  
Although the temperature differences induced by tropical cyclone passage
occur on relatively small temporal scales, there is a nontrivial seasonal
effect experienced by the ocean during the tropical cyclone season, at
the time scales considered in our analysis.  We focus here on the methodology
for estimating the seasonal cycle and fully detail the resulting empirical
findings
in Section~\ref{sect:application}.  In order to properly account for the
seasonal effect, we adapt previous methodology
\citep{ridgway_ocean_2002,roemmich_20042008_2009} to estimate a global mean
field specifically for
the ``counterfactual,'' i.e., we characterize the state of ocean temperatures
during the TC season, using the non-TC profiles of
Section~\ref{sub:tc_profiles_and_non_tc_profiles}.

Using the non-TC profiles of Section \ref{sub:tc_profiles_and_non_tc_profiles},
we estimate, separately for each pressure level $z = 10, 20, \dotsc, 200$ and
for the vertically averaged temperature values, a global mean field
through a variant of the Roemmich--Gilson climatology.  The Roemmich--Gilson
climatology \citep{roemmich_20042008_2009} was
among the first to take advantage of the spatial coverage of the Argo array to
characterize the mean state and annual cycle of temperature in the global upper
ocean.  Specifically, they used a local linear regression model 
\citep{ridgway_ocean_2002} for estimating
the climatological mean temperature as a function of location $(\Lon, \Lat)$,
pressure level $\Pres$,
and day of year $\YD$. This model is written down explicitly in
\cite{kuusela_locally_2018}, and our model,
which differs marginally from that of
\cite{roemmich_20042008_2009} in that we do not weight observations based on
distance from the center point $(\Lon^*, \Lat^*)$, nor do we use data from
pressure levels $\Pres' \neq \Pres$ to form estimates at pressure level
$\Pres$,
is elucidated in Equation~\eqref{eq:mean-field-estimation}.

At each of these twenty pressure levels (or for the vertical average) and at
each gridpoint of a mesh consisting of
180 equally spaced latitude
points and 360 equally spaced longitude points, we fit
\begin{align}\label{eq:mean-field-estimation}
    \begin{split}
        m(\Lon, \Lat, \YD) 
        =  &\beta_0 + [\text{
            first- and second-order terms in $\Lat$ and $\Lon$
        }]  \\
        &+ \sum_{k=1}^6 \gamma_k \sin \left(
            2\pi k\frac{\YD}{365.25}
        \right)
        + \sum_{k=1}^6 \delta_k \cos \left(
            2\pi k\frac{\YD}{365.25}
        \right)
    \end{split},
\end{align}
to the temperature observations,
using all the non-TC profiles that fall within
$\pm\SeasonalWindowSize^\circ$ (longitude/latitude) of the $(\Lon^*, \Lat^*)$
point at which our local regression function is centered.
Since these models are fit separately over each window and depth, the
estimation procedure can be trivially parallelized.

Let us denote by $m_{\Lon^*, \Lat^*}$ the mean function derived by
fitting a local linear regression with data within a window centered at 
$(\Lon^*, \Lat^*)$ using \eqref{eq:mean-field-estimation} for a fixed depth
$z$ or for the vertically averaged temperature values.
Then, for each of the TC profiles
(Section~\ref{sub:tc_profiles_and_non_tc_profiles}),
we produce the seasonal mean-adjusted temperature value in the following
fashion.
For a profile located at $x = (\Lon, \Lat)$ at time $\YD$ with temperature
value $\TempProf$, obtain the closest grid
point at which a local linear function was fitted: 
\begin{equation}
    (\Lon', \Lat') = \arg\min_{x'} \norm{x' - x}_2^2.
\end{equation}
Then, at each pressure level $\Pres = 10, 20, \dotsc, 200$, we obtain the
mean-adjusted profiles:
\begin{equation}\label{eq:seasonally-corrected-temperature}
    \SATempProf = \TempProf
        - m_{\Lon', \Lat'}(\Lon', \Lat', \YD).
\end{equation}
For each temperature value $\TempProf$ in the TC profile set, we apply this
mean adjustment to obtain the seasonally adjusted temperature value
$\SATempProf$.  In particular, for a \TextBefore{}-\TextAfter{} temperature
pair $\RawTempBefore, \RawTempAfter$, occurring at times
$\TimeBefore, \TimeAfter$, we obtain the seasonally adjusted temperature
value pair $\TempBefore, \TempAfter$.
This gives rise to the \emph{seasonally adjusted temperature differences}
\begin{equation}\label{eq:seasonally-adjusted-temp-diffs}
    \TempDiff = \TempAfter - \TempBefore.
\end{equation}
In terms of the high-level model \eqref{eq:high-level-model},
Equation~\eqref{eq:seasonally-adjusted-temp-diffs} in fact represents
the temperature difference without the seasonal effect.  This may be
seen through the decomposition:
\begin{align*}
    \TempDiff
    &= \TempAfter - \TempBefore \\
    &= \RawTempAfter
        - m_{\Lon', \Lat'}(\Lon', \Lat', \YDAfter)
        - (\RawTempBefore
        - m_{\Lon', \Lat'}(\Lon', \Lat', \YDBefore))\\
    &= (\RawTempAfter - \RawTempBefore)
    - (m_{\Lon', \Lat'}(\Lon', \Lat', \YDAfter)
    - m_{\Lon', \Lat'}(\Lon', \Lat', \YDBefore))\\
    &= \text{Temperature difference} - \text{Seasonal effect}   \\
    &\triangleq\text{Seasonally adjusted temperature difference}.
\end{align*}
since the last two mean field terms only differ in terms of the contribution
of the seasonal harmonics in Equation~\eqref{eq:mean-field-estimation}.

\begin{figure}
	\begin{subfigure}{.46\textwidth}
		\centering
		\includegraphics[width=1.\linewidth]{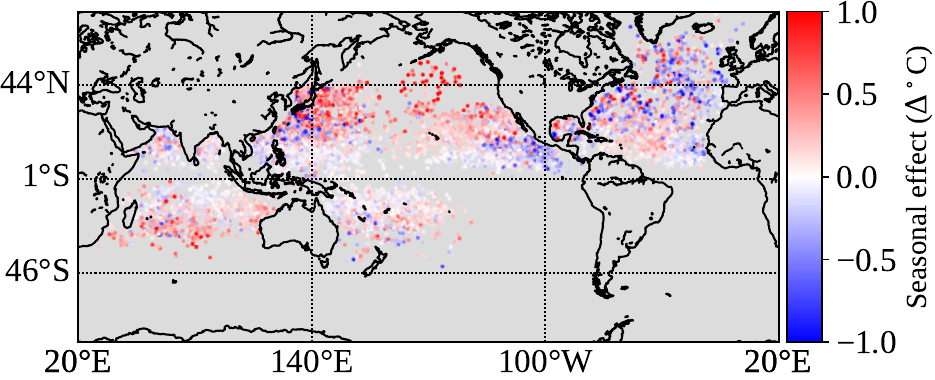}  
    \caption{$m_{{\xx}'}({\xx}', \TimeAfter)
      - m_{{\xx}'}({\xx'}, \TimeBefore)$}
        \label{fig:global-mean-field-fits}
	\end{subfigure}
  \hspace{0.2cm}
	\begin{subfigure}{.43\textwidth}
		\centering
		\includegraphics[width=1.\linewidth]{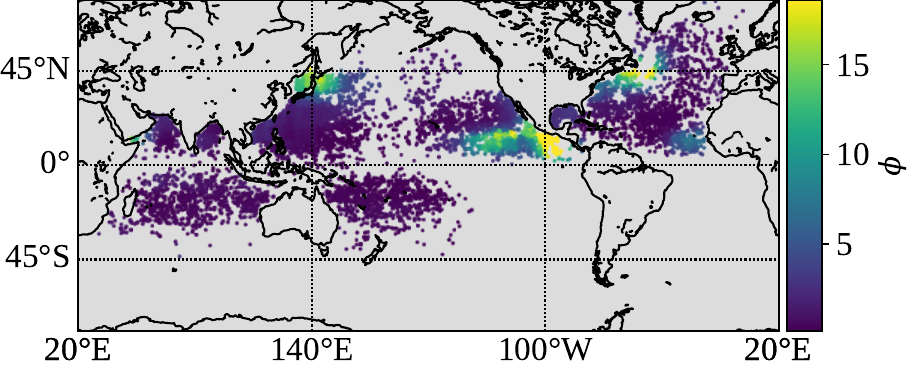}  
		\caption{Fitted $\phi$ parameters}
    \label{fig:global-variance-fits}
	\end{subfigure}
    \caption{
      The seasonal effects fitted to the signal profiles, displayed in
      Figure~\ref{fig:global-mean-field-fits}, show a general warming pattern
      during the TC season.
      The $\phi$ parameters, which are the dominant term of the ocean
      variability estimates, reveal in Figure~\ref{fig:global-variance-fits}
      higher noise in regions corresponding to
      the equatorial currents off Central America;
      Kuroshio and Oyashio currents off Japan;
      and Gulf Stream current off Nova Scotia
      \citep{talley_descriptive_2011}.
      Estimates at TC profile locations, pressure level $\Pres=40$
      are displayed here, though the underlying model fits are performed
      on a global grid.  The full set of estimates
      across all pressure levels deferred to the supplement
      \citep{spatiotemporal_tc_methods_supplement}.
    }
	\label{fig:fitted-mean-variance}
\end{figure}

Having accounted for the non-stochastic seasonal term, we may now directly
analyze the reduced model.  This is informally described as
\begin{equation}
  \label{eq:seasonal-reduced-high-level-model}
    \text{Seasonally adjusted temperature difference}
    = \text{TC signal} + \text{Ocean variability},
\end{equation}
and formally described, in terms of the ANOVA decomposition, as
\begin{equation}
  \label{eq:seasonal-reduced-fanova-model}
    T_\text{\TextAfter{}}
    -T_\text{\TextBefore{}}
    -(m_\text{\TextAfter{}}
    -m_\text{\TextBefore{}})
    = s(\XTrack, \DT)
    + (\VarAfterSmall 
      -\VarBeforeSmall),
\end{equation}
where we have economized the notation from
Equations~\eqref{eq:tempbefore-formal-decomp} and
\eqref{eq:tempafter-formal-decomp} for brevity.
The fitted seasonal means, at a depth of $\Pres = 40$, are depicted in
Figure~\ref{fig:global-mean-field-fits}, with the full set of estimates
across all pressure levels deferred to
supplementary material \citep{spatiotemporal_tc_methods_supplement}.  The
practical effect of this
seasonal adjustment is illustrated in Section~\ref{sect:app-mean-field}.
The mean field adjustment is also performed for the non-TC profiles,
yielding mean zero observations for the purposes of fitting the Gaussian
process model of Section \ref{sect:gaussian-process}.

\subsection{Variability estimates}\label{sect:gaussian-process}
Having obtained seasonally adjusted temperature differences at all pressure
levels, we next detail our method for estimating the variability of these
temperature differences.  Our general approach is an adaptation of the Gaussian
process model proposed by \cite{kuusela_locally_2018}, which handles the
globally
nonstationary covariance structure of the ocean using locally stationary
Gaussian processes, in a moving-window fashion
\citep{haas_lognormal_1990,haas_local_1995}. 
As in Section~\ref{sect:mean-field}, we learn the parameters of our model
using the \emph{non-TC} profiles, in order to capture the local dynamics of
the ocean without TC-induced effects.  However, once fitted, the model is used
to provide information about ocean variability in the time and locations
of the \emph{TC} profiles.  This is a subtle, but important, distinction.

Our local Gaussian process models are fitted on an equally spaced grid of 180
latitude points and 360
longitude points, with a time window centered on September.
This prioritizes
fidelity towards ocean dynamics during the late summer, early autumn
months when tropical
cyclones are prevalent in the Northern Hemisphere, where the majority
of the TC profile pairs are observed.
As before, this analysis is carried out separately at each pressure level
$\Pres = 10, \dotsc, 200$ and for the vertically averaged temperature
values.

In the following exposition, we follow the notation of
\cite{kuusela_locally_2018}, with slight adaptations to maintain
consistency with Section~\ref{sect:mean-field}.  We use the
term \emph{mean-adjusted} interchangeably with \emph{seasonally adjusted}
to emphasize that the mean-adjusted temperature values are
zero-mean (in the absence of TC signal), as required by the Gaussian
process model assumptions.

Let $\SATempProf$ denote a mean-adjusted temperature value associated with a
non-TC Argo profile at a
fixed pressure level.  At each
grid point $(\Lon^*, \Lat^*)$ and fixed pressure level $\Pres$, we assume the
following model for
data points $\SATempProf$ falling spatially within a window of
$\pm\GPWindowSize^\circ$ of the grid
point $(\Lon^*, \Lat^*)$ and temporally within the 3 months centered on
September:
\begin{equation}\label{eq:gaussian-process-model}
    \SATempProf_{i, j} = f_i(\xx_{i, j}, \YD_{i,j}) + \varepsilon_{i, j},
    \quad
    f_i \iid \GP(0, k(\xx_1, t_1, \xx_2, t_2; \GPParams)),
\end{equation}
where $i = 1, \dotsc, n$ indexes the years and $j = 1, \dotsc, m_i$
indexes the observations within the geospatial window for each year,
$\xx_{i,j}$ is the spatial location associated with the
mean-adjusted temperature value
$\SATempProf_{i, j}$, and $\YD_{i, j}$ is the time at which it
was observed.
The $\varepsilon_{i, j}$ is a Gaussian nugget effect
\citep{cressie_statistics_spatial_1993}, with 
$\varepsilon_{i,j} \iid N(0, \sigma^2)$, and
$\GP(0, k(\xx_1, \YD_1, \xx_2, \YD_2; \GPParams))$
denotes a zero-mean Gaussian process with a stationary
covariance function with parameters
$\GPParams = (\GPPhi, \GPThetaLat, \GPThetaLon, \GPThetaT)$.  The
specific function used is an anisotropic geospatial covariance function:
\begin{equation}
  \label{eq:gp-covariance-function}
    k(\xx_1, \YD_1, \xx_2, \YD_2; \GPParams)
    =
    \GPPhi\exp\{ - d(\xx_1, \YD_1, \xx_2, \YD_2)\}
\end{equation}
with $\GPPhi > 0$, where
\begin{equation}
  \label{eq:gp-distance-function}
    d(\xx_1, \YD_1, \xx_2, \YD_2)
    =
    \sqrt{
        \left(
            \frac{
                x_{\mathrm{lat}, 1} - x_{\mathrm{lat}, 2}
            }{
                \theta_\mathrm{lat}
            }
        \right)^2
        +
        \left(
            \frac{
                x_{\mathrm{lon}, 1} - x_{\mathrm{lon}, 2}
            }{
                \theta_\mathrm{lon}
            }
        \right)^2
        +
        \left(
            \frac{
                \YD_1 - \YD_2
            }{
                \theta_\YD
            }
        \right)^2
    }
\end{equation}
with $\theta_\mathrm{lat}, \theta_\mathrm{lon}, \theta_t > 0$.
As in \cite{kuusela_locally_2018}, these parameters are obtained through
the method of maximum likelihood.
As with the local mean field, the estimation of these parameters may be done
in parallel across windows and pressure levels.

A few remarks are in order.  First, we note that the fitted parameters
$\GPPhi(\xx)$, $\GPThetaLat(\xx)$, $\GPThetaLon(\xx)$, $\GPThetaT(\xx)$,
$\GPSigma(\xx)$ are all functions of $\xx = (\Lon, \Lat)$, as we assume local
stationarity to capture
global non-stationarity.  At each location, the Gaussian process model used to
estimate these parameters is centered on September, as most TCs in our
database occur during this time period.  We then use these
parameters to model the variance structure of profile pairs in all ocean
basins; one could refine these estimates by
fitting a model centered about every month and using the model
whose central month is closest in time to each profile pair.  Finally, we
observe that by construction,
the profile pairs from Section~\ref{sect:pairing-process} are assumed to have
the same locations, i.e., $x_{\mathrm{lat}, 1} = x_{\mathrm{lat}, 2}$ and
$x_{\mathrm{lon}, 1} = x_{\mathrm{lon}, 2}$.  Therefore, the covariance between
the two profiles in a profile pair is given by the following reduced
expressions.
For a mean-adjusted temperature value pair $\TempBefore, \TempAfter$
located at
$(\Lon, \Lat)$ with times $\YDBefore, \YDAfter$, let $\xx' = (\Lon', \Lat')$
be the nearest grid point, as in Section~\ref{sect:mean-field}, and let
$\GPPhi, \GPThetaT, \GPSigma$ as in \eqref{eq:gaussian-process-model}:
\begin{align}\label{eq:prof-pair-cov}
    \var(\TempBefore)
    = \var(\TempAfter)
    &= \GPPhi(\xx') + (\sigma(\xx'))^2;\\
    \cov(\TempBefore, \TempAfter)
    &=
    \GPPhi(\xx')\exp\left\{
        -\left|
            \frac{
                \YDBefore - \YDAfter
            }{
                \GPThetaT(\xx')
            }
        \right|
    \right\}.
\end{align}

To give an impression of the general variability in the global
ocean, we plot the fitted $\GPPhi$ parameters,
for a fixed pressure level $\Pres = 40$,
in Figure~\ref{fig:global-variance-fits}.
The parameters $\GPPhi$ are plotted because $\GPPhi$ is the dominant
term in each variability estimate.
The full set of figures depicting the fitted parameters and
temperature difference variances, at each pressure level, is deferred to
Section 5 of the supplementary material
\citep{spatiotemporal_tc_methods_supplement}.

We note that even though only
$\GPPhi$, $\GPThetaT$, and $\GPSigma$ are used to evaluate the data
covariances, the full model (\eqref{eq:gaussian-process-model},
\eqref{eq:gp-covariance-function}, and \eqref{eq:gp-distance-function})
must be fitted to estimate these parameters, and we require the full
Argo database, including the non-TC profiles, to obtain these estimates.

For simplicity, we assume that two profiles which do not share the same
\TextBefore{}
profile are uncorrelated.
However, there are cases in which the lineage
of a \TextBefore{} profile includes multiple \TextAfter{} profiles.  In these
cases, we
account for the covariance between the profiles in the lineage.  We detail the
case in which there are two \TextAfter{} profiles with
mean-adjusted temperature values $\TempAfterJ, \TempAfterK$, associated
with a
\TextBefore{} profile with mean-adjusted temperature value
$\TempBeforeI$, but the following framework is trivially extended
to an arbitrary number of \TextAfter{} profiles in the lineage.  In practice,
there
will be no more than six \TextAfter{} profiles for a given \TextBefore{}
profile, due to the
separation constraint introduced in Section~\ref{sect:pairing-process}.
Define the shorthand
$\TempIJK = \bmat{\TempBeforeI & \TempAfterJ & \TempAfterK}^\top$. 
We suppress the location $\xx'$ at which the parameters
$\GPPhi, \GPThetaT, \GPSigma$ are evaluated, for brevity.  First, consider the
covariance structure between the mean-adjusted temperature values
of the three profiles in this lineage (one baseline
$\TempBeforeI$ and two signal $\TempAfterJ, \TempAfterK$ values, occurring at
times
$\YDBeforeI, \YDAfterJ, \YDAfterK$):
\begin{equation}\label{eq:multi-prof-pair-cov}
    \cov(\TempIJK) = \bmat{
        \GPPhi + \GPSigma^2     
            &   
        \GPPhi\exp\{-|\frac{\YDAfterJ - \YDBeforeI}{\GPThetaT}|\}
            &   
        \GPPhi\exp\{-|\frac{\YDAfterK - \YDBeforeI}{\GPThetaT}|\}
        \\
        \GPPhi\exp\{-|\frac{\YDBeforeI - \YDAfterJ}{\GPThetaT}|\}
            &   
        \GPPhi + \GPSigma^2     
            &   
        \GPPhi\exp\{-|\frac{\YDAfterK - \YDAfterJ}{\GPThetaT}|\}
        \\
        \GPPhi\exp\{-|\frac{\YDBeforeI - \YDAfterK}{\GPThetaT}|\}
            &   
        \GPPhi\exp\{-|\frac{\YDAfterJ - \YDAfterK}{\GPThetaT}|\}
            &   
        \GPPhi + \GPSigma^2     
    }.
\end{equation}
The covariance matrix for the temperature differences may then be obtained as
follows.  Define the mean-adjusted temperature differences
\begin{align}
  \TempDiffJ &= \TempAfterJ - \TempBeforeI  \\
  \TempDiffK &= \TempAfterK - \TempBeforeI
\end{align}
and the shorthand $\TempDiffJK = \bmat{\TempDiffJ & \TempDiffK}^\top$. 
Observe that $\TempDiffJK$ is a simply a linear transformation of $\TempIJK$:
\begin{equation}
    \TempDiffJK = \bmat{
        \TempDiffJ   \\
        \TempDiffK
    }   =
    \bmat{
        -1  &   1   &   0   \\
        -1  &   0   &   1
    }\bmat{
        \TempBeforeI \\
        \TempAfterJ \\
        \TempAfterK \\
    }.
\end{equation}
In general, the transformation matrix is a column of minus ones horizontally
concatenated with the identity matrix.  It immediately follows that:
\begin{equation}\label{eq:multi-pair-tempdiff-covariance}
    \cov(\TempDiffJK) = \bmat{
        -1  &   1   &   0   \\
        -1  &   0   &   1
      }\cov(
        \TempIJK
      )\bmat{
        -1  &   1   &   0   \\
        -1  &   0   &   1
    }^\top.
\end{equation}
Recall that because the temperature differences have been mean-adjusted,
they are assumed to be zero-mean in the absence of a TC 
signal.  Therefore, when there is no TC signal, the distribution
of $\TempDiffJK$ is Gaussian with
mean zero and covariance $\cov(\TempDiffJK)$, given in
\eqref{eq:multi-pair-tempdiff-covariance}.  
Mathematically speaking, we have accounted for the
$(\VarAfterSmall - \VarBeforeSmall)$
variability term of \eqref{eq:seasonal-reduced-fanova-model}.
This forms the basis for the estimation of $\TCSignal$, our key quantity
of interest, in Section~\ref{sect:thin-plate-splines}.

\subsubsection{Computational remarks}
\label{ssub:computational_remarks}
Denote by $\TempDiffAll$ the full set of mean-adjusted temperature
differences, ordered such that profiles in the same lineage are
contiguous.  Its covariance, $\cov(\TempDiffAll)$,  is a block diagonal
matrix, with dense blocks corresponding to the lineages of distinct
\TextBefore{} profiles.
This is important for two reasons.  First, a block diagonal matrix admits an
efficient inverse; namely, the inverse of a block diagonal matrix may be
obtained by inverting the blocks individually.  Each of these blocks is no
more than $6$ entries wide, due to the separation constraint placed on profiles
in Section~\ref{sect:pairing-process}.  Second, the sparse structure
of the block diagonal matrix may be exploited for efficient linear algebra
operations \citep{2020SciPy-NMeth} necessary to construct the thin-plate
splines described in the sequel.

\subsection{Thin-plate splines}\label{sect:thin-plate-splines}
Having learned a model for the $(\VarAfterSmall - \VarBeforeSmall)$ term of
\eqref{eq:seasonal-reduced-fanova-model} through the Gaussian
process model of Section~\ref{sect:gaussian-process}, we find ourselves
properly positioned to estimate the TC-induced signal $s(\XTrack, \DT)$.
This we do with a thin-plate spline, a flexible nonparametric
technique for fitting a surface in two dimensions, subject to a curvature
penalty and observation weights.  Specifically,
we smooth the seasonally adjusted temperature differences using a fixed-knot
thin-plate spline
smoother
\citep{duchon1977splines,wahba1980spline,wahba_spline_1990,green_nonparametric_1994,
nychka_spatial_nodate,wood_generalized_nodate},
in which
we use the covariances estimated in Section~\ref{sect:gaussian-process}
to account for heteroskedasticity and cross-observation dependence
\citep{rao1973linear,draper1998applied,nychka_spatial_nodate,ruppert_semiparametric_2003}.
As with the local mean field model of
Section~\ref{sect:mean-field} and the Gaussian process model of
Section~\ref{sect:gaussian-process}, this procedure is conducted
separately at each pressure level $z \in \{10, 20, \dotsc, 200\}$
and for the vertically averaged profiles.

We use a fixed-knot smoother (rather than placing a knot at
every design point) chiefly in order to improve the conditioning of the linear
operators.  As an additional, computational consideration, the use of fixed
knots reduces the time and space complexity of the procedure.
The fixed knots are equally spaced in the $[-8, +8] \times [-2, +20]$
domain, in a grid formed from the cross product of 33 points in the cross-track
axis and 45 points in the time
axis.  This yields a knot every 0.5 units along each dimension, including on
the boundary.

The weighting of observations by the covariance between temperature differences
is also noteworthy.  From a data model standpoint, it is known that we are
using observations with drastically different noise levels $\VarBefore$,
$\VarAfter$, due to the effect of time and location.  Since we have
estimated these noise levels in Section~\ref{sect:gaussian-process},
the covariance-reweighting allows us to directly incorporate this
information when estimating $s(\XTrack, \DT)$ in the reduced model
\eqref{eq:seasonal-reduced-fanova-model}.
From a practical standpoint, the covariance-reweighting has a real
effect in improving the signal-to-noise ratio, as shown in Figure
\ref{fig:tps-main} of Section~\ref{sect:app-tps}.
The full details of the thin-plate spline smoother, including its motivation
and derivation, are given in Appendix~\ref{app:thinplatespline}.

Let $\tx_1, \dotsc, \tx_n \in \Omega \subset \RR^2$ be the design points, and
$y_1, \dotsc, y_n \in \RR$ be the corresponding responses.  We further fix a
grid of
regularly-spaced knots,
$\tilde \tx_1, \dotsc, \tilde \tx_m \in \Omega$, $m \ll n$,
at which the basis functions of our smoother will be centered.  We use
the shorthand notation
$\tX = \begin{bmatrix}\tx_1 \dots \tx_n\end{bmatrix}^\top$,
$\tilde\tX =
\begin{bmatrix}\tilde \tx_1 \dots \tilde \tx_n\end{bmatrix}^\top$,
and $\tY = \begin{bmatrix} \ty_1 \dots \ty_n\end{bmatrix}^\top$.
Finally, we denote by $\WW \in \RR^{n\times n}$ a weight matrix.
Traditionally, this is the inverse covariance matrix of the
observations $\tY$.

In our usage, we take $\tx_1, \dotsc, \tx_n$ to be the $\HurRefCoords$ coordinates
of our Argo profile pairs, as obtained from Section~\ref{sect:pairing-process};
$y_1, \dotsc, y_n$ to be the corresponding seasonally adjusted temperature
differences $\TempDiff$ of those pairs, as obtained from
Section~\ref{sect:mean-field};
and $\WW = \cov(\TempDiffAll)^{-1}$ to be the inverse matrix of covariances
between those seasonally adjusted temperature differences, as obtained from
Section~\ref{sect:gaussian-process}.
Therefore, here $\Omega = [-8, +8] \times [-2, +20]$, where the first
dimension (cross-track angle) is measured in degrees, and the second
dimension (time since TC passage) is measured in days.

The fixed-knot covariance-reweighted thin-plate spline problem is posed 
as follows.  We
seek coefficients $\TPSNFKnotCoefs, \TPSNFLinearCoefs$ such that the following
objective is minimized:
\begin{equation}\label{eq:tps-weighted-matrix-objective-methods}
    (\tY - \TPSFixedKnotBasis\TPSNFKnotCoefs
         - \LinBasisMatrix^\top\TPSNFLinearCoefs)^\top
    \WW
    (\tY - \TPSFixedKnotBasis\TPSNFKnotCoefs
         - \LinBasisMatrix^\top\TPSNFLinearCoefs)
    +
    \lambda \TPSNFKnotCoefs^\top
    \TPSFixedKnotSelfBasis
    \TPSNFKnotCoefs.
\end{equation}
The first term in \eqref{eq:tps-weighted-matrix-objective-methods} is a
generalized least squares
\citep{rao1973linear,draper1998applied,ruppert_semiparametric_2003}
data-fit criterion,
and the second term is a roughness
penalty \citep{meinguet_multivariate_1979,green_nonparametric_1994,
wood_generalized_nodate}, with
$\TPSFixedKnotBasis, \LinBasisMatrix, \TPSFixedKnotSelfBasis$
defined in the sequel.  The parameter $\lambda > 0$ mediates the competing
demands of the data fidelity and roughness penalty terms.

\subsubsection{Construction of the smoother}
The fixed-knot thin-plate spline smoother with covariance reweighting is
constructed as follows.  Denote by $\TPSFixedKnotBasis \in \RR^{n\times m}$ the
radial basis matrix with entries:
\begin{equation}\label{eq:fixed-knot-tps-basis-matrix}
    \TPSFixedKnotBasis_{ij} = \begin{cases}
        \frac{1}{16\pi}\norm{\tx_i - \tilde \tx_j}_2^2
            \log(\norm{\tx_i - \tilde \tx_j}_2^2)
        &\text{ for } \norm{\tx_i - \tilde \tx_j}_2^2 > 0   \\
        0 &\text{ otherwise }
    \end{cases}
\end{equation}
Notice that this consists of radial basis functions centered on the
knots $\tilde\tx_j$, evaluated at the data points $\tx_i$.  Further denote by
$\LinBasisMatrix \in \RR^{3\times n}$ the basis
matrix corresponding to the linear and constant terms:
\begin{align}
  \label{eq:tps-linear-basis-terms}
    \LinBasisMatrix = \begin{bmatrix}
        1       &       1       &       \cdots      &       1       \\
        \txi_{11}  & \txi_{21}  & \cdots    & \txi_{n1}  \\
        \txi_{12}  & \txi_{22}  & \cdots    & \txi_{n2}  \\
    \end{bmatrix}.
\end{align}
Finally, we define an additional ``basis'' matrix, where we evaluate the
radial basis functions at the knot locations $\tilde \tx_j$:
\begin{equation}\label{eq:fixed-knot-tps-self-basis-matrix}
    \TPSFixedKnotSelfBasis_{ij} = \begin{cases}
        \frac{1}{16\pi}\norm{\tilde\tx_i - \tilde\tx_j}_2^2
            \log(\norm{\tilde\tx_i - \tilde\tx_j}_2^2)
        &\text{ for } \norm{\tilde\tx_i - \tilde\tx_j}_2^2 > 0   \\
        0 &\text{ otherwise }
    \end{cases}
\end{equation}
for the purpose of evaluating the roughness penalty in
\eqref{eq:tps-weighted-matrix-objective-methods}.

The coefficients minimizing \eqref{eq:tps-weighted-matrix-objective-methods}
are given (cf. Appendix~\ref{app:thinplatespline}) by:
\begin{align}\label{eq:fixed-knot-thin-plane-spline}
    \begin{bmatrix}
        \TPSKnotCoefs       \\
        \TPSLinearCoefs
    \end{bmatrix}
    =
    \begin{bmatrix}
        \TPSFixedKnotBasis^\top \WW \TPSFixedKnotBasis  
            + \lambda\TPSFixedKnotSelfBasis
        &   \TPSFixedKnotBasis^\top\WW\LinBasisMatrix^\top  \\
        \LinBasisMatrix\WW\TPSFixedKnotBasis    
        &   \LinBasisMatrix\WW\LinBasisMatrix^\top
    \end{bmatrix}^{-1}
    \begin{bmatrix}
        \TPSFixedKnotBasis^\top\WW  \\
        \LinBasisMatrix\WW                      
    \end{bmatrix}
    \tY
\end{align}
where $\TPSKnotCoefs$ are the coefficients corresponding to the radial
basis functions centered at the fixed
knots and $\TPSLinearCoefs$ are the coefficients corresponding to the constant
and linear terms.  Estimates of the signal $\TCSignal$ at a new set of
locations $\hat \tx_1,
\dotsc, \hat \tx_N \in \RR^2$ may be obtained as follows:
\begin{align}
    \hat \tS =
    \begin{bmatrix}
        \TPSFixedKnotTestBasis      &   \LinFixedKnotTestBasis^\top
    \end{bmatrix}
    \begin{bmatrix}
        \TPSKnotCoefs       \\
        \TPSLinearCoefs
    \end{bmatrix}
\end{align}
where we denote $\TPSFixedKnotTestBasis\in\RR^{N\times m}$ with entries
\begin{align}
    \TPSFixedKnotTestBasis_{ij} = \begin{cases}
        \frac{1}{16\pi}\norm{\hat \tx_i - \tilde \tx_j}_2^2
            \log(\norm{\hat \tx_i - \tilde \tx_j}_2^2)
        &\text{ for } \norm{\hat\tx_i - \tilde \tx_j}_2^2 > 0   \\
        0 &\text{ otherwise }
    \end{cases}
\end{align}
and
\begin{align}
    \LinFixedKnotTestBasis = \begin{bmatrix}
        1       &       1       &       \cdots      &       1       \\
        \hat\txi_{11}  & \hat\txi_{21}  & \cdots    & \hat\txi_{N1}  \\
        \hat\txi_{12}  & \hat\txi_{22}  & \cdots    & \hat\txi_{N2}  \\
    \end{bmatrix}.
\end{align}

\subsubsection{Uncertainty quantification}
\label{ssub:tps-uncertainty-quantification}
Estimate variances and confidence intervals follow for the thin-plate
spline estimates through the linearity of the smoother and the
assumption of Gaussian noise.
Denote by $\TPSCov\in\RR^{n\times n}$ the observation covariance matrix
$\cov(\TempDiffAll)$ from \eqref{eq:multi-pair-tempdiff-covariance}.
The covariance matrix of the estimates is given by:
\begin{equation}
  \label{eq:estimate-covariance}
    \cov(\hat \tS)
    =
    \bmat{
        \TPSFixedKnotTestBasis  &   \LinFixedKnotTestBasis^\top
    }
    \AA^{-1}  
    \bmat{
        \TPSFixedKnotBasis^\top\WW  \\
        \LinBasisMatrix\WW                      \\
    } 
    \TPSCov
    \bmat{
        \TPSFixedKnotBasis^\top\WW  \\
        \LinBasisMatrix\WW                      \\
    }^\top
    \AA^{-1}
    \bmat{
        \TPSFixedKnotTestBasis  &   \LinFixedKnotTestBasis^\top
    }^\top
\end{equation}
where
\begin{align}
  \label{eq:tps-AA-shorthand}
    \AA = 
    \begin{bmatrix}
        \TPSFixedKnotBasis^\top \WW \TPSFixedKnotBasis  
            + \lambda\TPSFixedKnotSelfBasis
        &   \TPSFixedKnotBasis^\top\WW\LinBasisMatrix^\top  \\
        \LinBasisMatrix\WW\TPSFixedKnotBasis    
        &   \LinBasisMatrix\WW\LinBasisMatrix^\top
    \end{bmatrix}.
\end{align}
The form \eqref{eq:estimate-covariance} follows from the usual quadratic
form of covariances.

The matrix $\cov(\hat \tS) \in \RR^{N\times N}$ can be exorbitantly large
(e.g., in our use case, we produce estimates on a grid with $N= 400 \times 100
= 40000$ locations).  However, the construction of
pointwise confidence intervals (i.e.,
$\hat \ts_i \pm 1.96 \cdot \sqrt{\var(\hat\ts_i)}$,
appealing to the Gaussian assumption to obtain 95\% coverage)
require only evaluation of the diagonal entries of $\cov(\hat \tS)$, allowing
for significant computational savings.  We remark that although the confidence
intervals derived have width matching the standard deviation of the
estimates, the estimates $\hat \tS$ themselves are biased due to shrinkage;
therefore
the confidence intervals technically have biased centers.  It is thus possible
that the intervals are misplaced, or are too short, in regions where
$\TCSignal$ experiences high curvature.  We discuss alternatives for
improvement in Section~\ref{sect:discussion}.

\section{Estimated subsurface ocean thermal response to tropical cyclones}\label{sect:application}
\subsection{Seasonal mean field adjustment}
\label{sect:app-mean-field}
\begin{figure}[ht!]
	\begin{subfigure}{0\textwidth}
		\phantomcaption
		\label{fig:heat-flux-raw-diffs}
	\end{subfigure}
	\begin{subfigure}{1\textwidth}
		\centering
		\includegraphics[width=1.\linewidth]{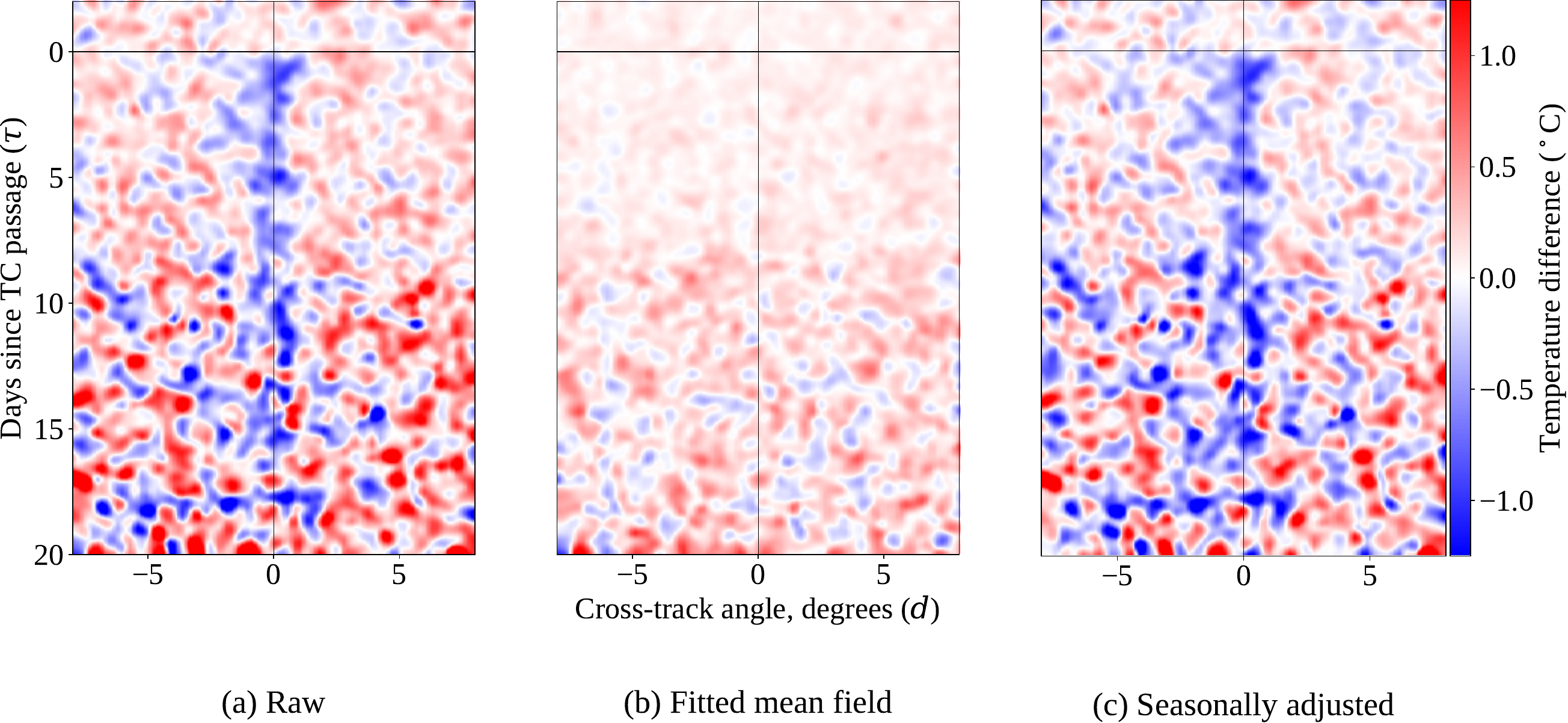}  
		\phantomcaption
		\label{fig:heat-flux-mean-field}
	\end{subfigure}
	\begin{subfigure}{0\textwidth}
		\phantomcaption
		\label{fig:heat-flux-corrected-diffs}
	\end{subfigure}
    \caption{
        Temperature differences in the $(\XTrack, \DT)$ coordinate
        system, at a depth of 40 meters.  Here we include all profile pairs
        across all storm intensities, and subject the data to a local constant
        smoother with an isotropic Gaussian kernel of bandwidth $\sigma=0.2$.
    }
	\label{fig:heat-flux-diffs}
\end{figure}

First, we detail the effect of the seasonal mean field adjustment, described
in Section~\ref{sect:mean-field}, on the observed data.  Recall that the model
\eqref{eq:mean-field-estimation} is fitted separately at each pressure level
$\Pres = 10, \dotsc, 200$.  In Figure~\ref{fig:heat-flux-diffs}, we depict
the effect of the seasonal adjustment on temperature differences at a pressure
level of $\Pres = 40$ dbar.  The raw temperature differences are displayed
in Figure~\ref{fig:heat-flux-raw-diffs}, with the seasonal differences plotted
in Figure~\ref{fig:heat-flux-mean-field}, and the seasonally adjusted
temperature differences in Figure~\ref{fig:heat-flux-corrected-diffs}.
The latter is the result of subtracting
Figure~\ref{fig:heat-flux-mean-field} from
Figure~\ref{fig:heat-flux-raw-diffs}.  The individual differences are subject
to
an isotropic Gaussian kernel smoother with a bandwidth of $\sigma=0.2$, in
order to more clearly display the data.

Figure~\ref{fig:heat-flux-mean-field} shows a warming effect that
increases in magnitude as the time between the signal profile and TC
passage increases.  This is consistent with the notion that the ocean
experiences an overall warming effect at this depth during the TC season.  Note
that
because we are plotting the difference in the mean field evaluated at the
times and locations of actual TC-associated profile pairs, we are only
examining the mean field differences during the tropical cyclone season of
each ocean basin analyzed.

The warming effect in the seasonal mean field difference is a consistent
feature in the pressure levels $\Pres=30$ through $\Pres=150$.  The pressure
levels $\Pres \in \{10, 20\}$ do not show a warming
pattern and might even exhibit a slight cooling effect,
while the pressure levels $z \geq 160$ show only a small contribution of the
seasonal signal to differences in paired profiles.
These phenomena are consistent with the spatial structure of the
seasonal cycle \citep{roemmich_20042008_2009}.
An adjustment for seasonal warming was not performed in a previous analysis
of ocean thermal response using paired Argo profiles 
\citep{cheng_global_2015}; as a result, some of the warming effects previously
observed over longer time spans may in fact have been due to seasonal variation.
The full sequence of results at all twenty pressure levels is deferred
to Section 1 of the supplementary material
\citep{spatiotemporal_tc_methods_supplement}.

\subsection{Ocean thermal response as a continuous function of time and cross-track distance}
\label{sect:app-tps}
\begin{figure}[ht!]
	\begin{subfigure}{0\textwidth}
		\centering
		\phantomcaption{
        }
		\label{fig:tps-no-cov}
	\end{subfigure}
	\begin{subfigure}{1\textwidth}
		\centering
		\includegraphics[width=1.\linewidth]{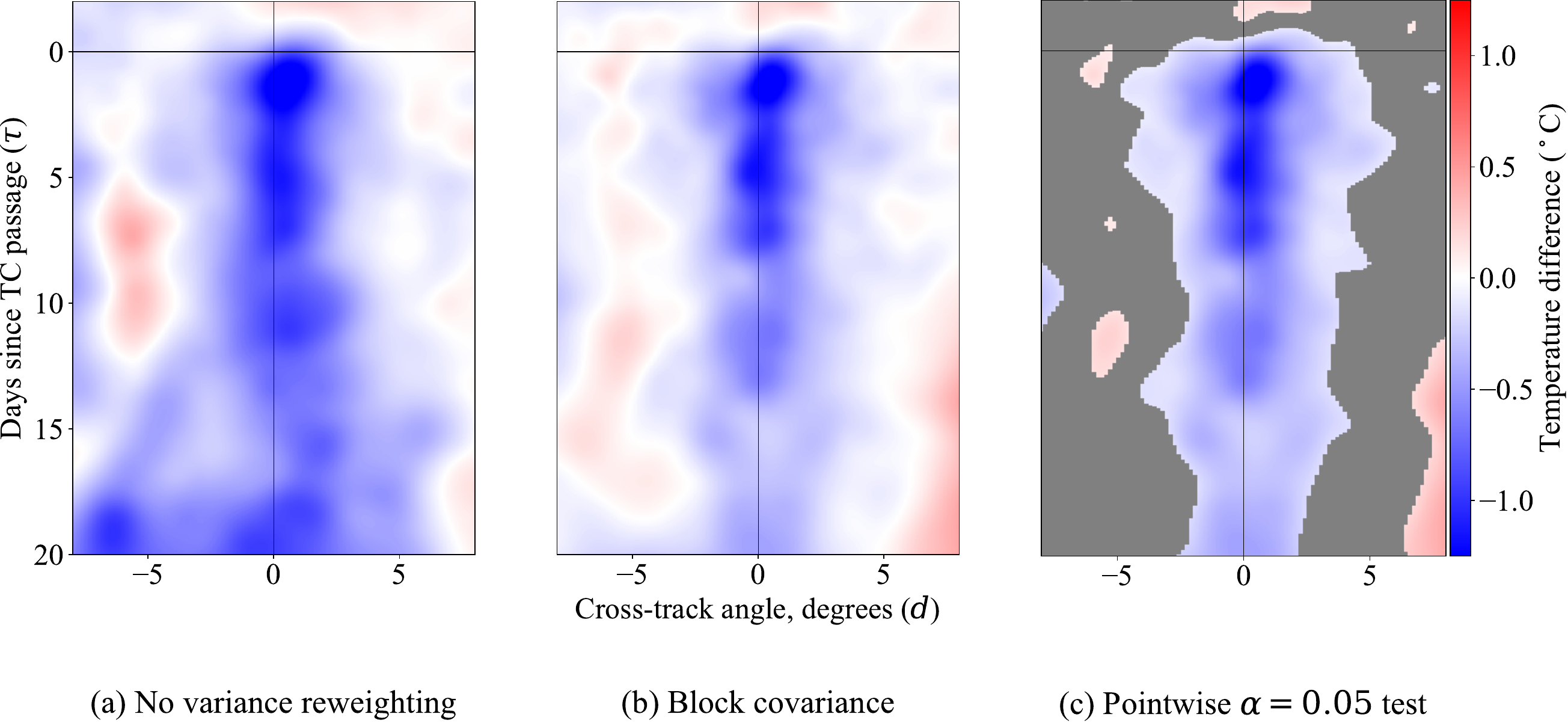}  
		\phantomcaption
    \label{fig:tps-block-cov}
	\end{subfigure}
	\begin{subfigure}{0\textwidth}
		\centering
		\phantomcaption
    \label{fig:tps-block-cov-masked}
	\end{subfigure}
    \caption{
        Temperature differences for hurricane-strength profile pairs in the
        $(\XTrack, \DT)$ coordinate system,
        at a depth of 10 meters, smoothed using thin-plate splines.
        We observe in Figure~\ref{fig:tps-no-cov} an
        unusual positive anomaly at cross-track $d\approx-6$, time difference
        $\DT \in (5, 13)$, as well as unusually strong negative differences
        at time difference $\DT\in(18, 20)$.
        Reweighting by the estimated block covariance in
        Figure~\ref{fig:tps-block-cov} attenuates these anomalies, and
        a pointwise hypothesis test in
        Figure~\ref{fig:tps-block-cov-masked} masks these phenomena,
        indicating that they were due to ocean variability noise unrelated
        to TCs.
    }
    \label{fig:tps-main}
\end{figure}

Having adjusted the temperature differences for the seasonal change using
the model detailed in Section~\ref{sect:mean-field} and accounted for the
variability across profile pairs via Section~\ref{sect:gaussian-process}, we
now apply the thin-plate
spline smoother to the seasonally adjusted temperature differences to obtain
our main scientific result: a characterization of the ocean thermal response
to the passage of hurricane-strength tropical cyclones as a continuous function
of time and cross-track distance.  This function is depicted for data from
profile pairs associated with a hurricane-strength tropical cyclone (formally
defined as having sustained wind speed of at least 64 knots at
$\TProj$, the time of TC passage) at a pressure level of $\Pres = 10$ decibars
in Figure~\ref{fig:tps-main}.

The thin-plate splines are fit separately at each depth $z$.  Within each
$z \in \{10, 20, \dotsc, 200\}$, we select a tuning parameter
$\hat\lambda^\text{cv}_z$ guided by leave-one-out cross-validation (LOOCV).  
Importantly, we do not take the $\hat\lambda^\text{min}$ that directly
minimizes the LOOCV score.  Instead, we take the $\lambda <
\hat\lambda^\text{min}$ that produces a LOOCV error slightly (1\%)
larger than the minimum LOOCV error.  This is done because heterogeneity
in the thermal response signal causes a direct minimization of the LOOCV score
to oversmooth the signal.  
The full details of the cross-validation procedure, including its mathematical
statement, plots illustrating the estimated test error achieved
by different $\lambda$, and the chosen $\hat\lambda^\text{cv}_z$, are deferred
to Appendix~\ref{app-subsect:leave_one_out_cross_validation}.

The central Figure~\ref{fig:tps-block-cov} depicts the thin-plate spline fit,
accounting for correlation between observations, at a depth of $z=10$ decibars.
Formally, the correlation adjustment is performed as follows.  We obtain an
estimate for $\TPSCov = \cov(\TempDiffAll)$ through the Gaussian process model
of Section~\ref{sect:gaussian-process}, and provide this to our
covariance-reweighted
thin-plate spline smoother of Section~\ref{sect:thin-plate-splines} by
setting $\WW = \TPSCov^{-1}$.  Recall that by construction, $\TPSCov$ is a
block diagonal matrix, with blocks no larger than six entries wide; therefore,
$\TPSCov^{-1}$ may be efficiently computed.  For a very small number of
temperature differences, the fitted variance is extremely small, on the order of
$\leq \exp\{-4.5\}$.  We truncate the left tail and filter out these
observations to prevent distorting the fit.  This omits only 488 observation
pairs out of 16025.

\begin{figure}[ht!]
	\begin{subfigure}{0\textwidth}
		\phantomcaption
    \label{fig:tps-cooling}
	\end{subfigure}
	\begin{subfigure}{\textwidth}
		\centering
		\includegraphics[width=1.\linewidth]{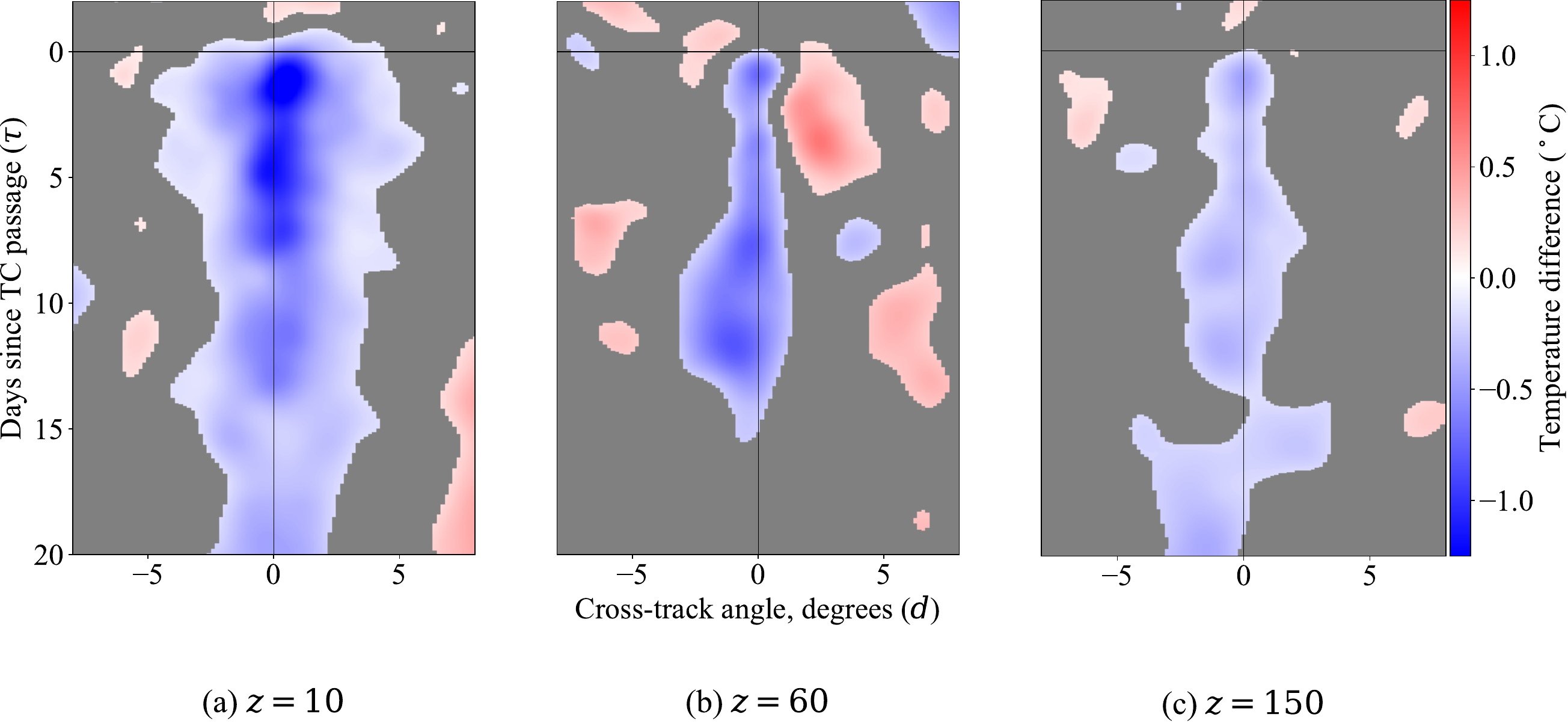}  
		\phantomcaption
    \label{fig:tps-block-cov-vertical-mixing}
	\end{subfigure}
	\begin{subfigure}{0\textwidth}
		\phantomcaption
		\label{fig:tps-stable}
	\end{subfigure}
    \caption{
        Temperature differences for hurricane-strength profile pairs, at depths
        of
        $\Pres = 10, 60, 150$, smoothed using thin-plate splines.
        Vertical mixing gives rise to a warming phenomenon on the right side
        of the tropical cyclone at the intermediate depths.  All plots are
        masked with a pointwise level $\alpha=0.05$ hypothesis test.
    }
    \label{fig:tps-vertical-mixing}
\end{figure}

In Figure~\ref{fig:tps-block-cov-masked}, we perform a pointwise
level $\alpha=0.05$ hypothesis test on the thin-plate spline fit, using
the variance estimates derived in
Section~\ref{sect:thin-plate-splines}.  The issues of performing a pointwise
hypothesis
test with $N=40000$ points are manifold, and we defer discussion of extensions
through which
we could make the significance test more sophisticated to
Section~\ref{sect:discussion}.  Nonetheless, we see in
Figure~\ref{fig:tps-block-cov-masked} a simple fact: that the estimated cooling
effect
induced by tropical cyclones is of a greater magnitude than the variability
of the estimates.  Moreover, this cooling effect persists for a sustained
period of time, stretching into two weeks following the tropical cyclone
passage.  We also see a rightward bias in the cooling effect, which may be
attributed to the storm's direction of rotation.\footnote{
    Due to the Coriolis effect, TCs in the Northern Hemisphere
    will induce a greater cooling effect on the right side of the
    cross-track angle parameterization than on the left side, due to the
    combined effects of forward movement and counterclockwise wind rotation.
    TCs in the Southern Hemisphere, which experience clockwise wind rotation,
    will induce a greater cooling effect on the left side.
    In order to display this effect
    in a consistent manner across hemispheres, we flip the sign of the
    cross-track angle for TCs in the Southern Hemisphere when producing
    Figures~\ref{fig:heat-flux-diffs}--\ref{fig:integrated-threepanel}.
}

To illustrate the importance of estimating the correlations between observations
on the final thin-plate spline fit, we plot in
Figure~\ref{fig:tps-no-cov} the surface estimated by applying
a thin-plate spline smoother without accounting for the variability estimates
from Section~\ref{sect:gaussian-process}, i.e., setting $\WW$ to be the
identity matrix.  These estimates are tuned with the same LOOCV procedure.  We
observe a sustained warm anomaly at $d\approx-6$ that is present over several
days.  Concerningly, this warm anomaly is comparable
in magnitude with the expected cooling effect observed at $d \approx 0$.
We also see negative anomalies which occur late in the recovery stage, between
$\DT\in(18, 20)$.
When viewed in context with Figures \ref{fig:tps-block-cov} and
\ref{fig:tps-block-cov-masked}, we see that adjusting for the correlations
reduces the impact of some outlying observations causing these anomalous
effects in the naive estimate, and the pointwise hypothesis test masks away
the effects almost entirely.

An additional phenomenon of scientific interest that may be observed
in the thin-plate spline fits is that of \emph{vertical mixing}, in which
the warm water close to the ocean surface is forced downwards by the
force of the tropical cyclone \citep{mei_spatial_2013}.  Although the net
change in heat content incidental to the passage of a tropical cyclone is
negative, vertical mixing may register as a warming effect at shallower
subsurface ocean depths.
This phenomenon may be observed in Figure~\ref{fig:tps-vertical-mixing}.
In Figure~\ref{fig:tps-cooling} we recall the smoothed temperature differences
at depth $\Pres=10$, in which a dramatic cooling effect is depicted.  At the
deeper pressure level of $\Pres=60$, however, we see a mixture of both cooling
and warming; namely, a net increase in temperature is observed at
$d\in(2, 4)$, $\DT\in (0, 6)$.  Finally, we see in Figure~\ref{fig:tps-stable}
that by $\Pres = 150$, the warming effect has disappeared,
while the cooling effect remains.  We note that the
characterization of this warming effect as a continuous function of both
cross-track angle and time since TC passage is made possible by our methodological
framework and was not possible in previous literature
\citep{cheng_global_2015}.

The discussion of vertical mixing is further developed in
Sections~\ref{sect:app-depth-time}, \ref{sect:app-depth-cross-track}, and
\ref{sub:thin_plate_spline_estimate_isosurface}.
The full set of covariance-reweighted thin-plate spline fits, at all pressure
levels $\Pres = 10, \dotsc, 200$, is
deferred to Section 2 of the supplementary material 
\citep{spatiotemporal_tc_methods_supplement}.

\subsection{Depth-time parameterization}
\label{sect:app-depth-time}
\begin{figure}[ht!]
	\begin{subfigure}{0\textwidth}
    \phantomcaption
		\label{fig:depth-time-d-2}
	\end{subfigure}
	\begin{subfigure}{1\textwidth}
		\centering
		\includegraphics[width=1.\linewidth]{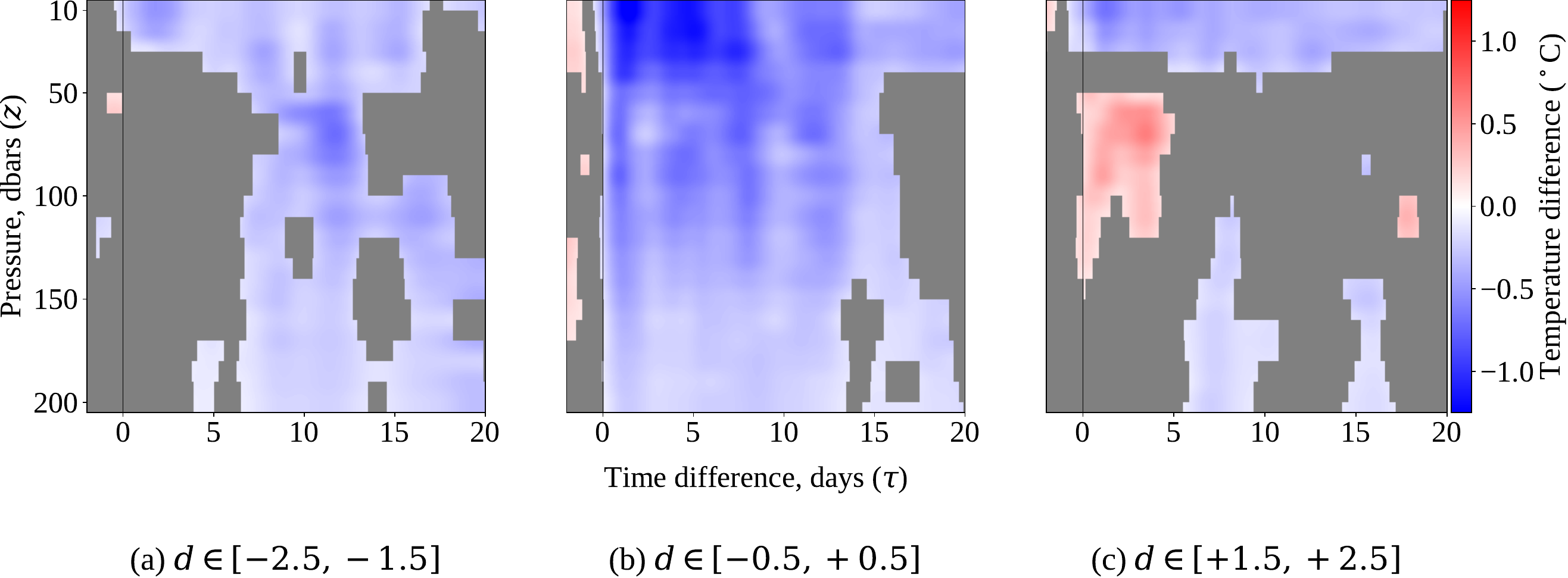}  
    \phantomcaption
		\label{fig:depth-time-d0}
	\end{subfigure}
	\begin{subfigure}{0\textwidth}
    \phantomcaption
		\label{fig:depth-time-d2}
	\end{subfigure}
    \caption{
        The covariance-reweighted thin-plate spline fits (hurricane-strength
        TCs) of
        Section~\ref{sect:app-tps} are averaged along three sets of cross-track
        angles $d$ corresponding to the left, center, and right relative to
        the TC's forward movement.
        The resulting estimates are plotted as a function of time since TC
        passage $\DT$ and
        pressure level $\Pres$.  A pointwise $\alpha=0.05$ hypothesis test is
        performed and used to mask areas where we fail to
        reject the null hypothesis.
    }
    \label{fig:depth-time}
\end{figure}

Of additional scientific interest is the characterization of ocean thermal
response for a fixed cross-track angle, as a function of time and pressure.
To facilitate this analysis, we collect the twenty thin-plate spline fits
for pressure levels $\Pres = 10, \dotsc, 200$ from Section~\ref{sect:app-tps}
and ``marginalize'' them along three sets of cross-track angles:
$d \in [-2.5, -1.5]$, $d \in [-0.5, +0.5]$, and $d \in [+1.5, +2.5]$.
These three plots are presented in Figure~\ref{fig:depth-time}.
In each of the marginalized plots, we subset the covariance-reweighted thin-plate
spline fits from Section~\ref{sect:app-tps} to each of the cross-track bins and
average out the cross-track component.  Variance estimates follow immediately
(owing to the linearity of averaging) and are used to construct a pointwise
$\alpha=0.05$ hypothesis test.  We display these marginalized estimates
in Figure~\ref{fig:depth-time}.

The cross-track angle bins, centered on $d = -2, 0, +2$, were chosen to
emphasize three particular oceanographic and meteorological phenomena.
First,
the bin centered on $d=0$ in Figure~\ref{fig:depth-time-d0} depicts the ocean cooling
induced by the passage of tropical cyclones.  Of particular note is the
fact that this cooling appears to persist for a sustained period of time,
specifically, up to at least two weeks.  As one may expect, the cooling effect
diminishes
as pressure $\Pres$ or time since TC passage $\DT$ increases.
Second, the bin centered at $d=+2$ in Figure~\ref{fig:depth-time-d2} depicts the
phenomenon of vertical mixing previously discussed in
Section~\ref{sect:app-tps}.  
As TC winds mix surface water downward, we observe warming at depths between
$\Pres = 50$ to $\Pres = 150$.
Past $\Pres = 150$, we find
that the warming effect has faded away.  Finally, we contrast the strong
warming effect observed in Figure~\ref{fig:depth-time-d2} with the lack of
warming (at a statistically significant level) in
Figure~\ref{fig:depth-time-d-2}, which depicts the bin centered on $d=-2$.
This
reflects the fact that the effect of vertical mixing is asymmetric, due to the
counterclockwise rotation of tropical cyclones in the Northern
Hemisphere.

\subsection{Depth-cross-track parameterization}
\label{sect:app-depth-cross-track}
We further demonstrate that our continuous-time
methodology captures the temporally binned results of \cite{cheng_global_2015},
whose profile pairing process we used in this paper.  To introduce temporal
binning, we take the covariance-reweighted thin-plate spline estimates from
Section~\ref{sect:app-tps} across pressure levels $\Pres = 10, \dotsc, 200$
and split them into two bins: $\DT\in[0, +3)$ and $\DT\in[+3, +20)$, termed
the ``forced'' stage and ``recovery'' stage by \cite{cheng_global_2015}.
For each of the two bins, we perform the same averaging process
as in Section~\ref{sect:app-depth-time} to marginalize the estimates over the
$\tau$ axis and obtain $\alpha=0.05$ pointwise significance tests.
\begin{figure}[ht!]
	\begin{subfigure}{1\textwidth}
		\centering
		\includegraphics[width=1.\linewidth]{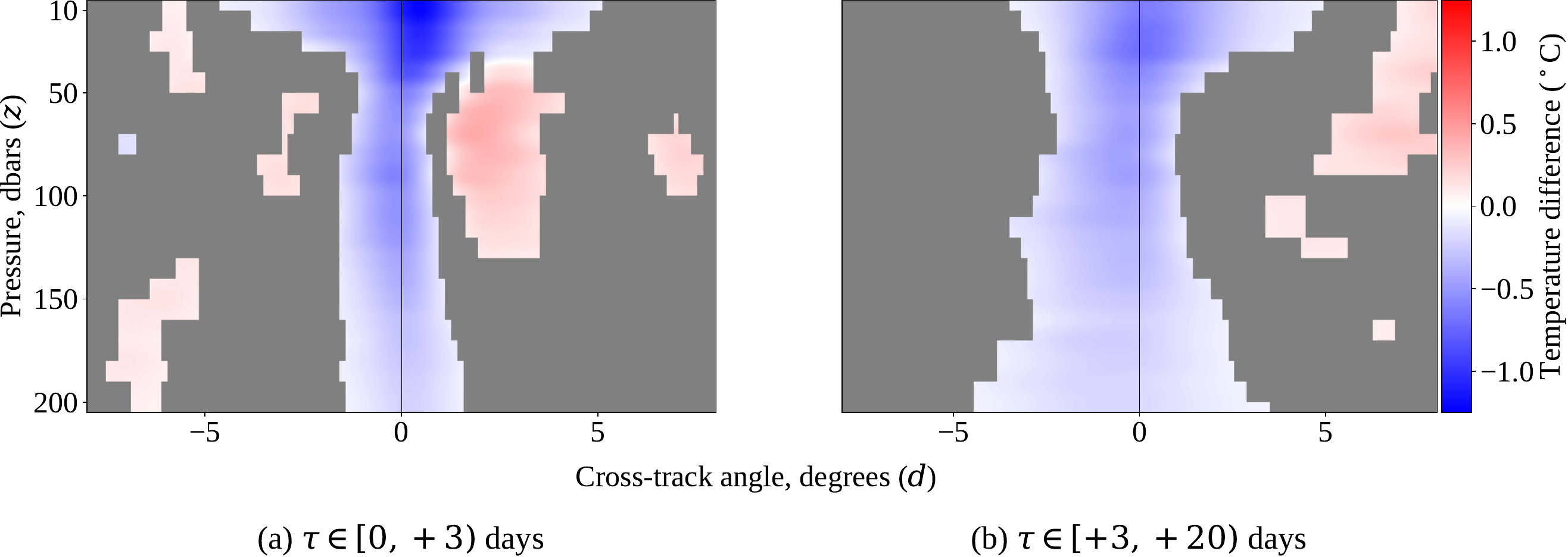}  
    \phantomcaption
		\label{fig:depth-crosstrack-forced}
	\end{subfigure}
	\begin{subfigure}{0\textwidth}
    \phantomcaption
		\label{fig:depth-crosstrack-recovery}
	\end{subfigure}
    \caption{
        The covariance-reweighted thin-plate spline fits (hurricane-strength
        TCs) of
        Section~\ref{sect:app-tps} are averaged along two sets of time since TC
        passage
        $\tau$, and the resulting estimates are plotted as a function of
        pressure level $\Pres$ and cross-track angle $\XTrack$.  A
        pointwise $\alpha=0.05$
        hypothesis test is performed and used to mask areas where we fail to
        reject the null hypothesis.
    }
    \label{fig:depth-crosstrack}
\end{figure}
We present
the resulting fits as functions of cross-track angle $\XTrack$ and pressure
level $\Pres$ in Figure~\ref{fig:depth-crosstrack}.

Several noteworthy phenomena are present in Figure~\ref{fig:depth-crosstrack}.
In Figure~\ref{fig:depth-crosstrack-forced}, we observe both the primary
cooling effect and the secondary warming effect in the subsurface on the right side of the storm
due to vertical mixing, as discussed in Sections~\ref{sect:app-tps}
and \ref{sect:app-depth-time}.  We also observe a sustained cooling effect in
Figure~\ref{fig:depth-crosstrack-recovery}, during the period called the
``recovery'' stage in \cite{cheng_global_2015}.
While the primary cooling effect observed in the two time periods is qualitatively
consistent with the results presented by \cite{cheng_global_2015}, their
results also show a strong warming effect on both sides of the storm during the
recovery stage (Figure 10 in \citealp{cheng_global_2015}), which is absent from
our estimate. Overall, it appears that the recovery stage estimate in
\cite{cheng_global_2015} is ``warm biased'' in comparison to our estimate. 
We attribute the difference to our methodology's accounting of the seasonal
mean field and variability estimates, which, when used in the thin-plate spline
estimate, attenuate the influence of outlying observations dominated by non-TC
oceanographic variability as illustrated in Figure~\ref{fig:tps-main}.

\subsection{Isosurfaces of the thin-plate spline estimates}
\label{sub:thin_plate_spline_estimate_isosurface}

\begin{figure}[ht!]
	\begin{subfigure}{0.32\textwidth}
		\centering
    \vspace{0.25cm}
		\includegraphics[width=1.\linewidth]{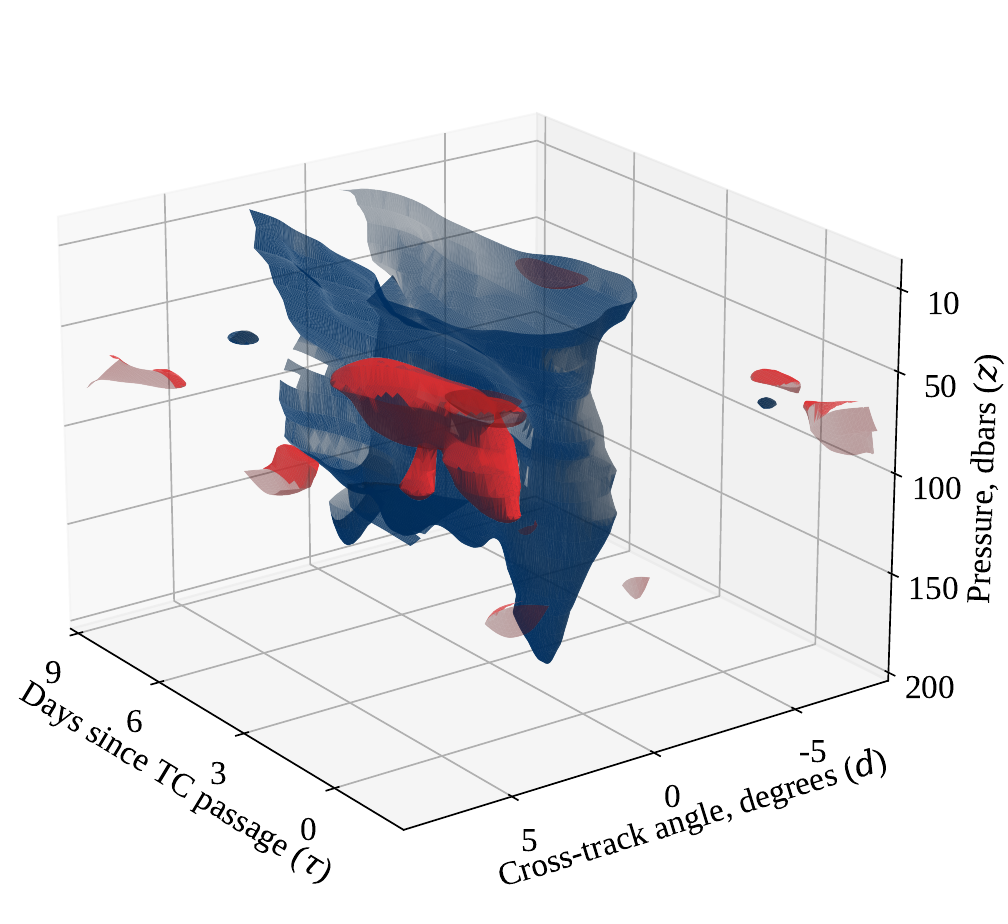}  
    \caption{
      $(\theta, \varphi) = (21.0^\circ, 144.1^\circ)$
    }
		\label{fig:isosurface1}
	\end{subfigure}
	\begin{subfigure}{0.32\textwidth}
		\centering
		\includegraphics[width=1.\linewidth]{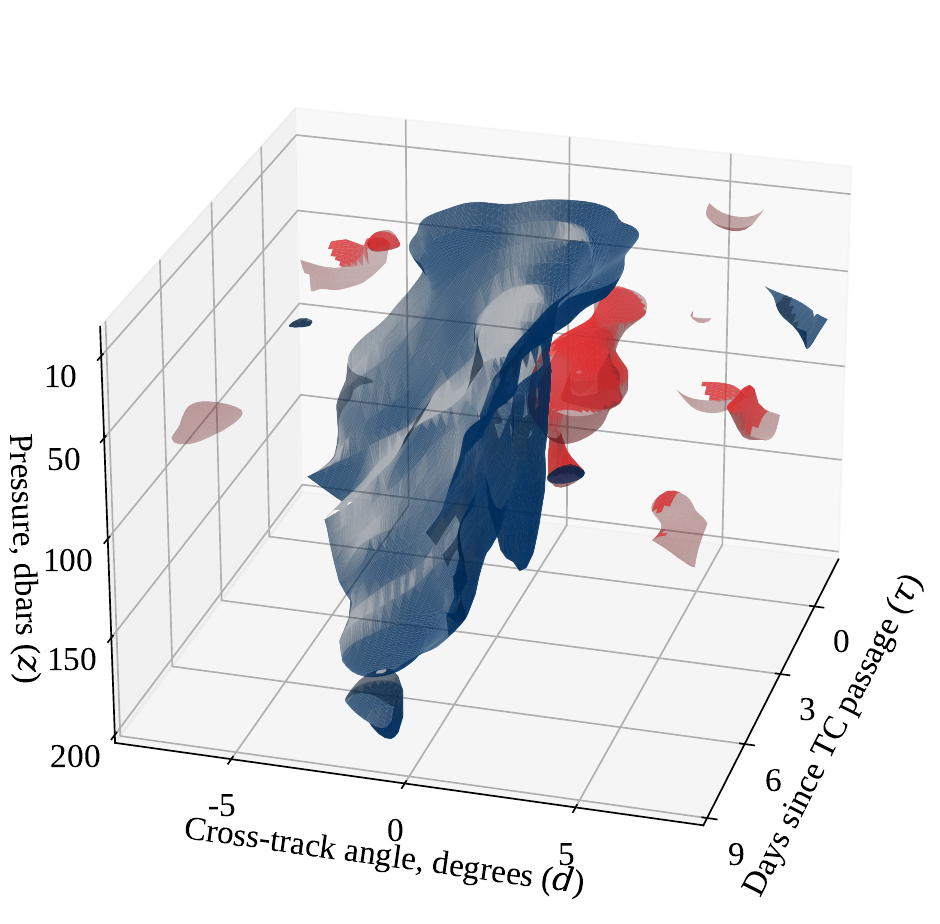}  
    \caption{
      $(\theta, \varphi) = (25.1^\circ, 16.1^\circ)$
    }
		\label{fig:isosurface2}
	\end{subfigure}
	\begin{subfigure}{0.32\textwidth}
		\centering
		\includegraphics[width=1.\linewidth]{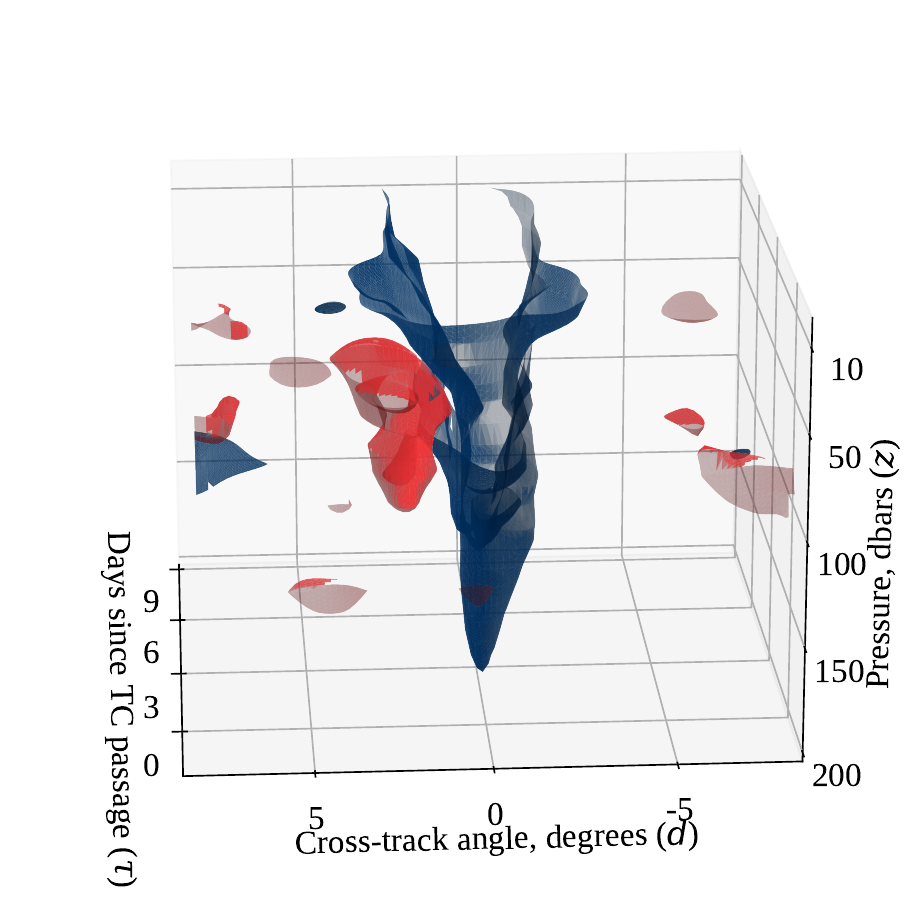}  
		\caption{
      $(\theta, \varphi) = (18.6^\circ, 176.4^\circ)$
    }
		\label{fig:isosurface3}
	\end{subfigure}
  \caption{
    $\pm0.3^\circ\mathrm{C}$ isosurfaces of the thin-plate spline fits, where
    negative temperature differences are shaded in blue and positive
    temperature differences are shaded in red.  Each subplot is labeled with
    its camera perspective, given in its polar ($\theta$) and azimuthal
    ($\varphi$) angles.
  }
  \label{fig:tps-isosurface}
\end{figure}

To provide a holistic understanding of the ocean thermal response structure
as a function of depth, cross-track distance, and time since TC passage,
we compute isosurface plots of the thin-plate spline estimates from
Section~\ref{sect:thin-plate-splines}.  Two sets of isosurfaces are computed;
one set
for regions of negative temperature differences, and one set for regions of
positive temperature differences.  These are shaded blue and red, respectively,
in Figure~\ref{fig:tps-isosurface}.

For each set of isosurfaces, we first mask the thin-plate spline estimates
that fail to reject a pointwise level $\alpha = 0.05$ hypothesis test, using
the variance estimates from Section~\ref{ssub:tps-uncertainty-quantification}.
We then apply the marching cubes algorithm
\citep{lorensen_marching_1987,lewiner_efficient_2012,2020SciPy-NMeth}
to identify a positive and a negative isosurface of magnitude
$0.3^\circ\mathrm{C}$.  The positive isosurface is colored in red, and the
negative in blue.

We observe in Figure~\ref{fig:tps-isosurface} a large blue mass corresponding
to the TC-related ocean cooling.  This blue mass
is restricted to a relatively small range of cross-track angles near the
center, but persists in time.  This behavior is consistent with
Figure~\ref{fig:depth-time}.  We also observe a smaller red mass,
corresponding to the local warming effect associated with the vertical
mixing induced by the TC, as previously witnessed in
Figure~\ref{fig:tps-vertical-mixing} and discussed in
Section~\ref{sect:app-depth-time}.  Consistent with Figures
\ref{fig:tps-vertical-mixing}, \ref{fig:depth-time}, and
\ref{fig:depth-crosstrack}, this warming effect is only observed on the
positive side of the cross-track distance axis.  Figure~\ref{fig:isosurface3}
mirrors Figure~\ref{fig:depth-crosstrack-forced} almost
exactly, the only differences being the viewing angle and filtering of
very small values inherent in finding nonzero isosurfaces.

\subsection{Vertically averaged temperature differences}
\label{sub:integrated_temperature_differences}

Here, we present the results from the vertically averaged analysis, 
in which we estimate the full model \eqref{eq:tempbefore-formal-decomp} and
\eqref{eq:tempafter-formal-decomp} using the vertically averaged
temperature values from Section~\ref{ssub:pchip_interpolation}.
As before, we choose the smoothing parameter
$\hat\lambda^\text{cv}$ guided by leave-one-out cross-validation, with the full
details deferred to Section \ref{app-subsect:leave_one_out_cross_validation}.

\begin{figure}[ht!]
	\begin{subfigure}{0\textwidth}
		\phantomcaption
		\label{fig:integrated-seasonally-adjusted}
	\end{subfigure}
	\begin{subfigure}{1\textwidth}
    \centering
    \includegraphics[width=1.0\linewidth]{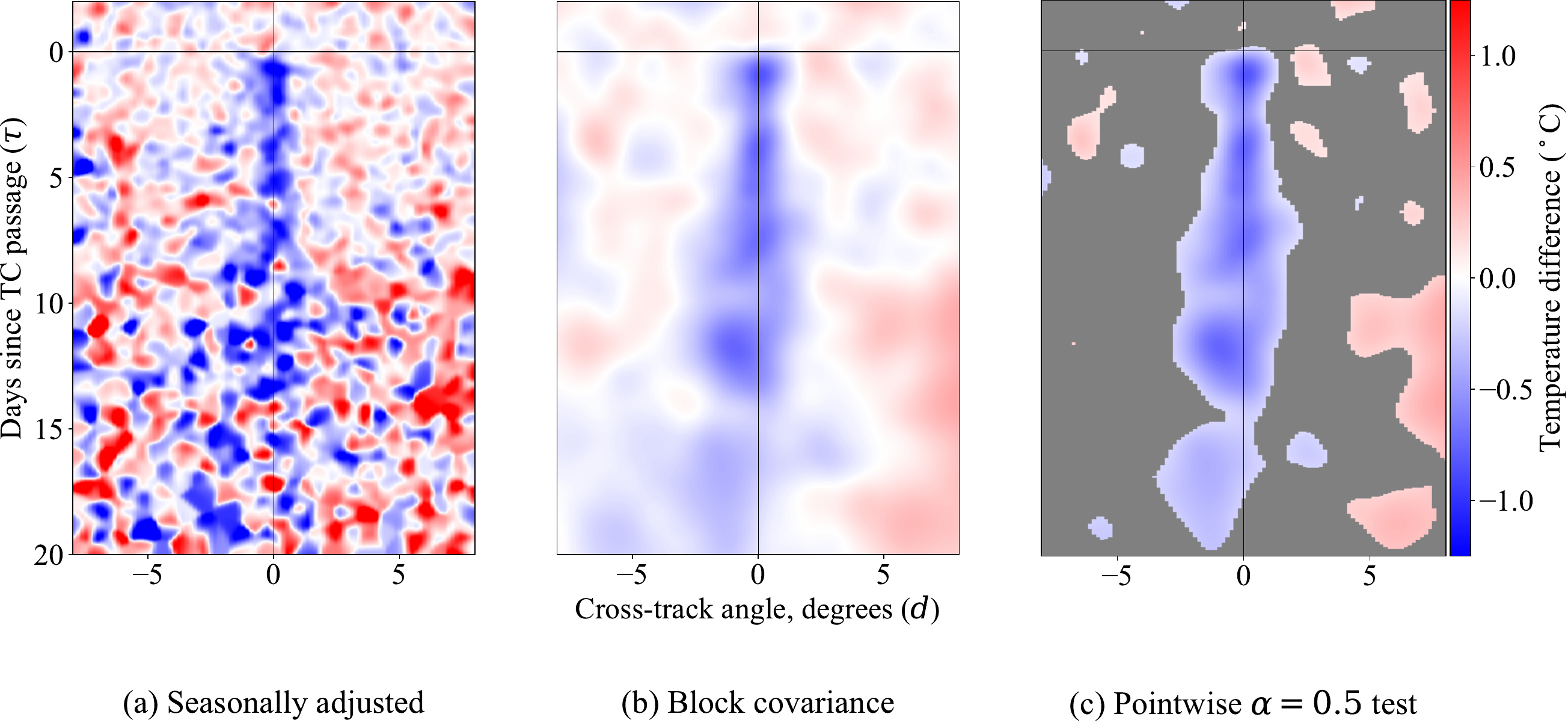}
		\phantomcaption
		\label{fig:integrated-tps-block-cov}
	\end{subfigure}
	\begin{subfigure}{0\textwidth}
		\phantomcaption
		\label{fig:integrated-tps-pointwise-test}
	\end{subfigure}
  \caption{
    Results from the vertically averaged analysis.
    The three subfigures depicted here are analogous to Figures
    \ref{fig:heat-flux-corrected-diffs},
    \ref{fig:tps-block-cov},
    and \ref{fig:tps-block-cov-masked}, respectively.
  }
  \label{fig:integrated-threepanel}
\end{figure}

The seasonally adjusted temperature differences and accompanying thin-plate
spline fits are presented in Figure~\ref{fig:integrated-threepanel}.  
In Figure~\ref{fig:integrated-seasonally-adjusted}, we only show data from
pairs associated with hurricane-strength TCs, in
contrast to Figure~\ref{fig:heat-flux-diffs}, which showed data from all
profile pairs regardless of storm intensity.  The restriction to
hurricane-strength profile pairs here is to provide a fairer comparison to the
thin-plate spline estimates of Figures \ref{fig:integrated-tps-block-cov} and
\ref{fig:integrated-tps-pointwise-test}, which use the hurricane-strength
subset (as in Figure~\ref{fig:tps-main}).

The thin-plate spline fit in Figure~\ref{fig:integrated-tps-block-cov} conforms
with our intuition; we observe a cooling signal, as in
Figure~\ref{fig:tps-block-cov}, but its magnitude is dampened by the
vertical averaging
procedure.  More importantly, Figure~\ref{fig:integrated-tps-pointwise-test}
does not show any warming effect indicative of vertical mixing. This is in
contrast to Figures~\ref{fig:tps-block-cov}, \ref{fig:depth-time-d2},
and \ref{fig:depth-crosstrack-forced}, which treat depth as a separate dimension
of analysis.  From an oceanographic and meteorological perspective, this
suggests that the vertical mixing-induced warming effect is local in depth; it
only appears at some pressure levels, because it is caused by the forcing of
warm surface waters downward by the TC.  Once we integrate vertically, the
local warming effect is cancelled out by a cooling effect near the surface and
is no longer visible.  From a statistical perspective, this underscores the
importance of treating depth as a separate axis of analysis, using the
vertical resolution of Argo observations.

\conclusions[Discussion]{
  \label{sect:discussion}
  In this paper, we have introduced a comprehensive methodological framework
based upon an ANOVA-type decomposition 
for analyzing tropical cyclone-induced ocean temperature changes.  This framework
includes a pairing process for identifying reference observations, the
construction of an annual mean field to account for seasonal shift at
the time scales of the measured differences, as well as estimation of
observation variability and computationally efficient smoothers that leverage
these estimates of variability.  

Aside from methodological contributions, we have presented several results
of scientific interest.  In particular, our framework improves upon past
scientific literature by modeling the tropical cyclone-induced ocean
thermal response
as a function of time, pressure, and cross-track angle, in which both the time
and cross-track angle components are treated continuously.  Through this
modeling, we are able to
produce fine-grained characterizations of the ocean thermal response at a
number of different depths, and we observe differing ocean thermal response
as a function of depth.  Specifically, we recover the primary effect of
surface cooling and secondary effect of subsurface warming due to vertical
mixing.  By computing isosurfaces on our thin-plate spline fits, we are
able to characterize these two phenomena as a function of the depth, time,
and cross-track distance dimensions.  Finally, the seasonal mean
field component of our methodological
framework reveals a nontrivial warming of the ocean near its surface during the
TC season, which we estimate and account for in our analysis.

Of course, this is not the full scientific story.  The ocean response to the
passage of TCs is complex, including contributions from upwelling near the
storm center, downwelling in the outer regions, and mixed layer entrainment
\citep{elsberry_mixed_1976,lin_ocean_2017};
horizontal currents \citep{dasaro_cold_2007}; in addition to air-sea exchanges
of heat \citep{emanuel_air-sea_1986,emanuel_thermodynamic_1999}. The initial
stratification in a region is also important for understanding the relevant
processes.
Nonetheless, the statistical methodology presented in this paper provides an
important new framework through which TC-induced temperature changes in the
upper
ocean may be studied as a function of depth, distance from TC track center,
and time since TC passage.

Although our methodological framework was developed
with ocean thermal response in mind, the techniques discussed are of general
interest.  An immediate generalization is the use of the exact same
methodology for Argo salinity profiles in lieu of temperature profiles.
One may expect that due to rainfall, evaporation, and mixing, the passage of a
tropical
cyclone would be registered in the salinity profiles of the affected ocean
region.  Similar analyses for other transient atmospheric
phenomena, including atmospheric rivers and southern ocean storm systems,
are also possible.  One could additionally regress these signals on latitude,
storm strength, pre-storm ocean state, ocean basin, and other covariates
to better model and understand the underlying scientific phenomena.
Regression on latitude may be particularly relevant with regard to the
ongoing scientific discourse on the large-scale climatological effects of
tropical cyclones
\citep{emanuel_contribution_2001,korty_tropical_2008,
jansen_impact_2009,jansen_seasonal_2010,haney_hurricane_2012}.

The present methodological framework also admits a number of possible
refinements.  From a statistical standpoint, one could improve upon our model by
considering each temperature profile as a functional observation and modeling
the
depth component continuously, rather than interpolating to a sequence of depths
and fitting separate models at each depth level.  This would potentially allow
one to also
infer changes in mixed layer depth and thermocline stability.  Our model
could also potentially be improved by replacing the thin-plate spline smoother
with
a spline smoother with a spatially varying smoothing parameter 
(e.g., \citealp{ruppert2000theory})
or a nonlinear, multiresolution smoother that naturally accounts for
heterogeneity in the TC-induced signal
(e.g., \citealp{silverman_wavelets_1999}).

The pointwise hypothesis tests used in Section~\ref{sect:app-tps} have a
number of limitations, including lack of correction for multiple testing and
smoothing bias.  These provide natural opportunities for improvement;
however, the problem of amending the pointwise hypothesis tests could be
subsumed by instead considering field significance 
(e.g., \citealp{livezey_statistical_1982, wilks_field_2006,
delsole_field_2011}).
Because the primary goal of this
paper is to introduce a general methodological framework of which the
final hypothesis tests are just one component, we defer these improvements
to future work.

An additional extension of statistical and scientific interest involves
carefully accounting for float drift.  In this paper, we require the baseline and
signal profiles to be within $0.2^\circ$ of each other and then assume that
they share a single location in the analysis, admitting a useful simplification
of the fitted covariance function.  An extension would involve relaxing the
strict closeness requirement of the paired profiles and removing the assumption
that the paired profiles are in the same location by more carefully modeling
their covariance, taking into account their relative spatial locations.  This
requires
a nontrivial generalization of the Gaussian process model of
\cite{kuusela_locally_2018}, but would provide the benefit of
increasing the number of available pairs with which to fit the model.  An
early approach in this vein is that of \citet{paciorek2006spatial}.  Yet
another extension would be the consideration of parametric models for
the TC wake \citep{haney_hurricane_2012}.  Finally, the fusion of
microwave sea surface temperature (mSST;
\citealp{gentemann_diurnal_nodate,wentz_three_2005})
observations with Argo
profiles might provide an opportunity to produce estimates on the
individual-storm level, rather than in terms of the global aggregates,
as was done here.
mSST observations are very dense in latitude and longitude, but lack the
depth penetration provided by the Argo profiles, which are relatively
sparse in latitude and longitude.  Therefore, mSST and Argo
are naturally complementary in the information provided.

}


\codeavailability{
  Code implementing our methodology may be found at
  \url{https://github.com/huisaddison/tc-ocean-methods}.
} 





\appendix
\section{Derivation of the thin-plate spline estimator}
\label{app:thinplatespline}
In this section, provided for completeness, we motivate and derive the
thin-plate spline estimator
\citep{duchon1977splines,wahba1980spline,wahba_spline_1990,green_nonparametric_1994,wood_thin_2003}
used in
Section~\ref{sect:thin-plate-splines} for completeness.
First, we review the natural thin-plate spline smoother in
Section~\ref{app-subsect:ntps-smoother}.
Sections~\ref{app-subsect:vr-tps-smoother} and
\ref{app-subsect:fk-tps-smoother} then
describe versions of the thin-plate spline when subject to covariance
reweighting (via generalized least squares;
\citealp{rao1973linear,draper1998applied,
nychka_spatial_nodate,ruppert_semiparametric_2003})
and when its basis functions are constructed from a fixed knot set
\citep{wahba1980spline,wood_generalized_nodate}.

\subsection{The natural thin-plate spline smoother}
\label{app-subsect:ntps-smoother}
Suppose we observe data $\tx_1, \dotsc, \tx_n \in \Omega \subset \RR^2$ with
responses $y_1, \dotsc, y_n \in \RR$.  Define the shorthand
notation $\tX = \begin{bmatrix}\tx_1 \dots \tx_n\end{bmatrix}^\top$
and $\tY = \begin{bmatrix} y_1 \dots y_n\end{bmatrix}$.

The natural thin-plate spline problem provides a smooth function $g$ such that
$g(\tx_i) \approx \ty_i$ for $i = 1, \dotsc, n$.  The ``closeness'' of $g(\tx_i)$
to $\ty_i$ is enforced through squared error, while a penalty on the curvature
of $g$ discourages roughness.  Formally, we define the roughness penalty:
\begin{align}\label{eq:tps-roughness-penalty}
    J(g)
    &=
    \iint_{\RR^2}
    \left(
        \frac{\partial^2 g}{\partial \txi_1^2}
    \right)^2
    +
    2\left(
        \frac{\partial^2 g}{\partial \txi_1\txi_2}
    \right)^2
    +
    \left(
        \frac{\partial^2 g}{\partial \txi_2^2}
    \right)^2
    \dint\txi_1 \dint\txi_2.
\end{align}
We then define an objective in which the regularization parameter $\lambda > 0$
mediates the competing demands of the squared error and roughness penalty
terms:
\begin{equation}\label{eq:tps-functional-objective}
    S(g)    =
    \sum_{i=1}^n
    (\ty_i - g(\tx_i))^2
    + 
    \lambda J(g).
\end{equation}
Let us define the basis matrix $\TPSBasisMatrix\in\RR^{n\times n}$, with
entries
\begin{equation}\label{eq:tps-basis-matrix}
    \TPSBasisMatrix_{ij} = \begin{cases}
        \frac{1}{16\pi}\norm{\tx_i - \tx_j}_2^2
            \log(\norm{\tx_i - \tx_j}_2^2)
        &\text{ for } \norm{\tx_i - \tx_j}_2^2 > 0   \\
        0 &\text{ otherwise }
    \end{cases}.
\end{equation}
One may recognize this as analogous to the basis matrix
\eqref{eq:fixed-knot-tps-basis-matrix} from
Section~\ref{sect:thin-plate-splines}, where we have replaced the fixed
knots
with a knot at each sample point.  Further recall
the basis of linear and constant terms \eqref{eq:tps-linear-basis-terms}.

It is possible
\citep{duchon1977splines,meinguet_multivariate_1979,wahba1980spline,
wahba_spline_1990,green_nonparametric_1994} to rewrite
\eqref{eq:tps-functional-objective} as
\begin{equation}\label{eq:tps-matrix-objective}
    S(g) =
    (\tY - \TPSBasisMatrix\TPSNFKnotCoefs
         - \LinBasisMatrix^\top\TPSNFLinearCoefs)^\top
    (\tY - \TPSBasisMatrix\TPSNFKnotCoefs
         - \LinBasisMatrix^\top\TPSNFLinearCoefs)
    +
    \lambda \TPSNFKnotCoefs^\top\TPSBasisMatrix\TPSNFKnotCoefs
\end{equation}
where $\TPSNFKnotCoefs\in\RR^n, \TPSNFLinearCoefs\in\RR^3$ uniquely define
the thin-plate spline $g$.  Expanding \eqref{eq:tps-matrix-objective},
we obtain:
\begin{align}
    S(g)
    &=  \bmat{
        \TPSNFKnotCoefs \\
        \TPSNFLinearCoefs
    }^\top \bmat{
        \TPSBasisMatrix^2   +   \lambda\TPSBasisMatrix
            &   \TPSBasisMatrix\LinBasisMatrix^\top \\
        \LinBasisMatrix\TPSBasisMatrix
            &   \LinBasisMatrix\LinBasisMatrix^\top
    }\bmat{
        \TPSNFKnotCoefs \\
        \TPSNFLinearCoefs
    }
    - 2\bmat{
        \TPSNFKnotCoefs \\
        \TPSNFLinearCoefs
    }^\top \bmat{
        \TPSBasisMatrix \\
        \LinBasisMatrix
    } \tY
    + \tY^\top \tY
\end{align}
Therefore, first-order optimality and convexity imply that
\eqref{eq:tps-basis-matrix} is minimized by $\hat\TPSNFKnotCoefs,
\hat\TPSNFLinearCoefs$ such that:
\begin{equation}\label{eq:tps-smoother-solution}
    \bmat{
        \hat\TPSNFKnotCoefs \\
        \hat\TPSNFLinearCoefs
    } = 
    \bmat{
        \TPSBasisMatrix^2   +   \lambda\TPSBasisMatrix
            &   \TPSBasisMatrix\LinBasisMatrix^\top \\
        \LinBasisMatrix\TPSBasisMatrix
            &   \LinBasisMatrix\LinBasisMatrix^\top
    }^{-1}\bmat{
        \TPSBasisMatrix \\
        \LinBasisMatrix
    } \tY
\end{equation}

\subsection{The covariance-reweighted thin-plate spline smoother}
\label{app-subsect:vr-tps-smoother}
Due to the correlated and heteroskedastic nature of the temperature differences
in this
paper, we desire a version of the thin-plate spline smoother
\eqref{eq:tps-smoother-solution} that properly accounts for the
these effects.  This is done via
generalized least squares \citep{rao1973linear,draper1998applied,
nychka_spatial_nodate,ruppert_semiparametric_2003}, which
poses the objective function:
\begin{equation}\label{eq:tps-weighted-matrix-objective}
    S(g) =
    (\tY - \TPSBasisMatrix\TPSNFKnotCoefs
         - \LinBasisMatrix^\top\TPSNFLinearCoefs)^\top
    \WW
    (\tY - \TPSBasisMatrix\TPSNFKnotCoefs
         - \LinBasisMatrix^\top\TPSNFLinearCoefs)
    +
    \lambda \TPSNFKnotCoefs^\top\TPSBasisMatrix\TPSNFKnotCoefs,
\end{equation}
where $\WW = \cov(\TempDiffAll)^{-1}$ is the inverse of the covariance matrix
between the seasonally adjusted temperature differences, previously estimated
in Section~\ref{sect:gaussian-process}.
In the simplest case of diagonal $\WW$,
\eqref{eq:tps-weighted-matrix-objective} may be interpreted as
inversely weighting the contribution of each observation to the objective
by the magnitude of its variance.

Carrying through an analogous quadratic expansion and minimization, we find
that \eqref{eq:tps-weighted-matrix-objective} is minimized by
$\hat\TPSNFKnotCoefs, \hat\TPSNFLinearCoefs$ such that:
\begin{equation}\label{eq:tps-weighted-smoother-solution}
    \bmat{
        \hat\TPSNFKnotCoefs \\
        \hat\TPSNFLinearCoefs
    } = 
    \bmat{
        \TPSBasisMatrix\WW\TPSBasisMatrix   +   \lambda\TPSBasisMatrix
            &   \TPSBasisMatrix\WW\LinBasisMatrix^\top \\
        \LinBasisMatrix\WW\TPSBasisMatrix
            &   \LinBasisMatrix\WW\LinBasisMatrix^\top
    }^{-1}\bmat{
        \TPSBasisMatrix\WW \\
        \LinBasisMatrix\WW
    } \yy.
\end{equation}

\subsection{Adaptation to fixed knots}
\label{app-subsect:fk-tps-smoother}
Finally, we detail the adaptation of the covariance-reweighted thin-plate
spline smoother to the fixed-knot regime
\citep{wahba1980spline,wahba_spline_1990,ruppert_semiparametric_2003,
wood_generalized_nodate}.
Suppose a set of fixed knots $\tilde \tx_1, \dotsc, \tilde \tx_m$, with
$\TPSFixedKnotBasis$ as defined in \eqref{eq:fixed-knot-tps-basis-matrix}
and $\TPSFixedKnotSelfBasis$ as defined in
\eqref{eq:fixed-knot-tps-self-basis-matrix}.
The thin-plate spline smoothing problem, restricted to this set of knots,
is given \citep{wood_generalized_nodate} by:
\begin{equation}\label{eq:tps-fixed-knot-weighted-matrix-objective}
    S(g) =
    (\tY - \TPSFixedKnotBasis\TPSNFKnotCoefs
         - \LinBasisMatrix^\top\TPSNFLinearCoefs)^\top
    \WW
    (\tY - \TPSFixedKnotBasis\TPSNFKnotCoefs
         - \LinBasisMatrix^\top\TPSNFLinearCoefs)
    +
    \lambda \TPSNFKnotCoefs^\top
    \TPSFixedKnotSelfBasis
    \TPSNFKnotCoefs
\end{equation}
where $\TPSNFKnotCoefs \in \RR^m$ and $\TPSNFLinearCoefs\in\RR^3$.  Carrying
through the same calculations,
we find that \eqref{eq:tps-fixed-knot-weighted-matrix-objective} is
minimized by $\hat\TPSNFKnotCoefs, \hat\TPSNFLinearCoefs$ such that:
\begin{equation}\label{eq:tps-fixed-knot-weighted-smoother-solution}
    \bmat{
        \hat\TPSNFKnotCoefs \\
        \hat\TPSNFLinearCoefs
    } = 
    \bmat{
        \TPSFixedKnotBasis^\top\WW\TPSFixedKnotBasis
        +   \lambda\TPSFixedKnotSelfBasis
            &   \TPSFixedKnotBasis^\top\WW\LinBasisMatrix^\top \\
        \LinBasisMatrix\WW\TPSFixedKnotBasis
            &   \LinBasisMatrix\WW\LinBasisMatrix^\top
    }^{-1}\bmat{
        \TPSFixedKnotBasis^\top\WW \\
        \LinBasisMatrix\WW
    } \tY.
\end{equation}

\section{Leave-one-out cross-validation}%
\label{app-subsect:leave_one_out_cross_validation}

In this section, we fully describe the procedure used to select the 
$\CVLamb_z$ in Section~\ref{sect:app-tps} and
$\CVLamb$ in Section~\ref{sub:integrated_temperature_differences}.  Informally,
we calculate the leave-one-out cross-validation (LOOCV) scores for a fine grid
of $\lambda$, and then choose the $\lambda$ corresponding to the most
complex model whose LOOCV score is within 1\% of the minimum LOOCV.
This procedure is motivated by the fact that the signal being estimated is
heterogeneous, and a naive minimization of the LOOCV error
was empirically found to result in an
over-regularized model.  This shall be made formal in the sequel.

The leave-one-out cross-validation score under weighted smoothing
\citep{cook_residuals_1982,green_nonparametric_1994} is given by
\begin{equation}
  \label{eq:loocv-weighted-smoothing-primitive}
  \mathrm{CV}(\lambda)
  = \sum_{i=1}^n w_i \left(\ty_i - \hat g^{(-i)}(\tx_i; \lambda)\right)^2,
\end{equation}
where $w_i$ is an observation-specific weight (typically the inverse
variance) and $\hat g^{(-i)}(\tx_i; \lambda)$ is the thin-plate spline,
fitted to all observations \emph{except} observation $i$, using
regularization parameter $\lambda$, evaluated on observation $i$.

As the thin-plate spline is a linear smoother, we may
trivially access the leave-one-out residuals through the
``hat matrix trick'' \citep{cook_residuals_1982,green_nonparametric_1994}
\begin{equation}
  \label{eq:hat-matrix-trick}
  \ty_i - \hat g^{(-i)}(\tx_i; \lambda)
  =
  \frac{
    \ty_i - \hat g(\tx_i; \lambda)
  }{
  1 - h^{(\lambda)}_{ii}
  },
\end{equation}
where $h^{(\lambda)}_{ii}$ is the $i$th diagonal entry of 
the ``hat matrix'' of $\hat g(\tx_i; \lambda)$,
i.e., the matrix $\HH^{(\lambda)}$ that satisfies $\HH^{(\lambda)} \yy =
\hat g(\tX; \lambda) = \hat \yy$.
Combining \eqref{eq:loocv-weighted-smoothing-primitive} and \eqref{eq:hat-matrix-trick},
we obtain the LOOCV score
\begin{equation}
  \label{eq:loocv-weighted-smoothing}
  \text{CV}(\lambda)
  = \sum_{i=1}^n
  \frac{1}{\var(\ty_i)}
  \left(
    \frac{
      \ty_i - \hat \ty_i
    }{
    1 - h^{(\lambda)}_{ii}
    }
  \right)^2
  ,
\end{equation}
where we have taken $w_i = \nicefrac{1}{\var(\ty_i)}$, as in
Section~\ref{sect:thin-plate-splines}.

\begin{figure}
	\begin{subfigure}{.49\textwidth}
		\centering
		\includegraphics[width=1.\linewidth]{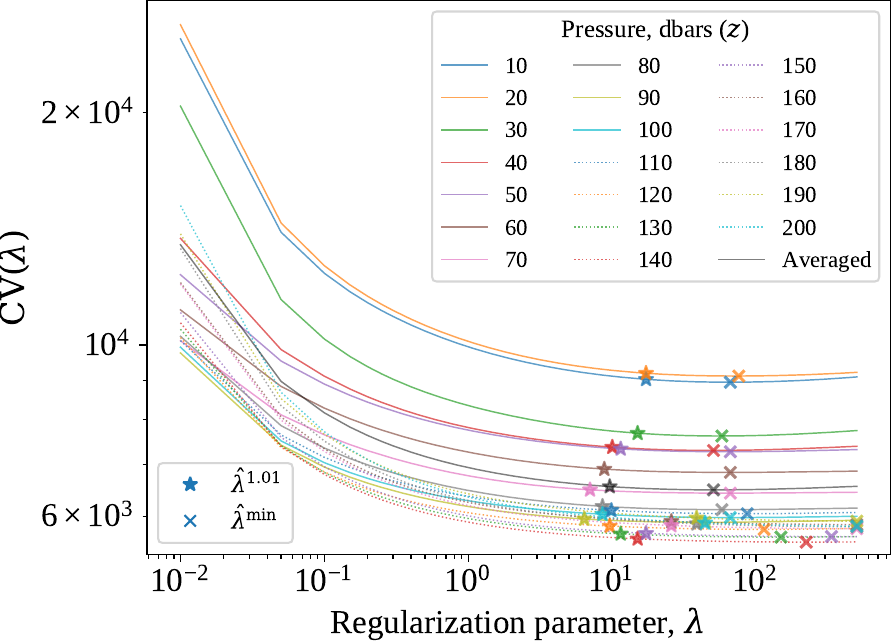}  
    \caption{
      Leave-one-out cross-validation errors
    }
		\label{fig:loocv-curves}
	\end{subfigure}
  \hspace{0.3cm}
	\begin{subfigure}{.46\textwidth}
		\centering
		\includegraphics[width=1.\linewidth]{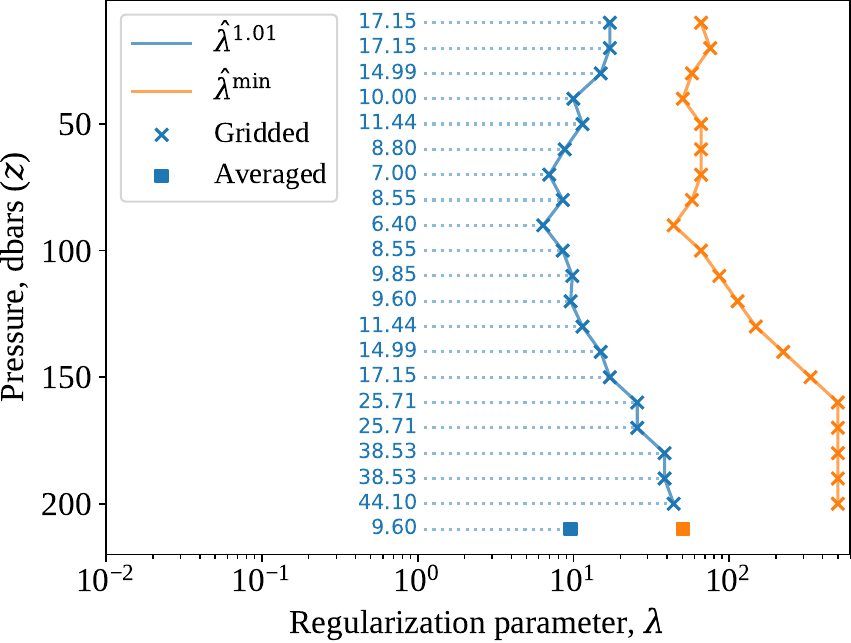}  
    \caption{
      Chosen $\lambda$ values
    }
    \label{fig:loocv-chosen-lambda}
	\end{subfigure}
    \caption{
      We observe Figure~\ref{fig:loocv-curves} that the minimizing
      $\hat\lambda^\text{min}$, for each depth, tends to lie in an extremely
      shallow basin,
      within which a wide range of $\lambda$ achieve comparable estimated
      test error (n.b., the vertical axis is on the logarithmic scale).  By
      choosing a $\hat\lambda^\text{1.01}$ associated with a 1\%
      increase in test error, we obtain fits that more faithfully illustrate
      the TC-induced thermal response.  Figure~\ref{fig:loocv-chosen-lambda}
      reports the chosen $\hat\lambda$ for each pressure level $z$ and for
      the vertically averaged temperature differences.  Fits using $\lambda >
      500$ are ignored as they essentially only fit the linear (unpenalized)
      terms of the thin-plate spline.
    }
	\label{fig:loocv-minimized-inflated}
\end{figure}

The LOOCV errors for $z \in \{10, 20, \dotsc, 200\}$ fits and for the
vertically averaged fit are illustrated in Figure~\ref{fig:loocv-curves},
with the minimizing $\hat\lambda^\text{min}$ are marked with crosses.  However,
as we alluded earlier, we do not use $\hat\lambda^\text{min}$; rather, we
take
\begin{equation}
  \label{eq:cv-lamb}
  \CVLamb = \hat\lambda^\text{1.01}
  = \min \{\lambda: \mathrm{CV}(\lambda) \leq
  1.01\cdot\mathrm{CV}(\hat\lambda^\text{min})\},
\end{equation}
which we mark with stars in Figure~\ref{fig:loocv-curves}.
This corresponds to taking the most complex model whose LOOCV
error is within 1\% of the minimized LOOCV error.  
We favor a more complex
model than the one obtained by naively minimizing the LOOCV error because
the signal being estimated is heterogeneous, i.e., it obeys different
levels of smoothness on different regions of the domain.  
The $\hat\lambda^\text{min}$ and $\hat\lambda^\text{1.01}$ are given as a
function of depth in  Figure~\ref{fig:loocv-chosen-lambda}.

\begin{figure}
	\begin{subfigure}{.48\textwidth}
		\centering
		\includegraphics[width=1.\linewidth]{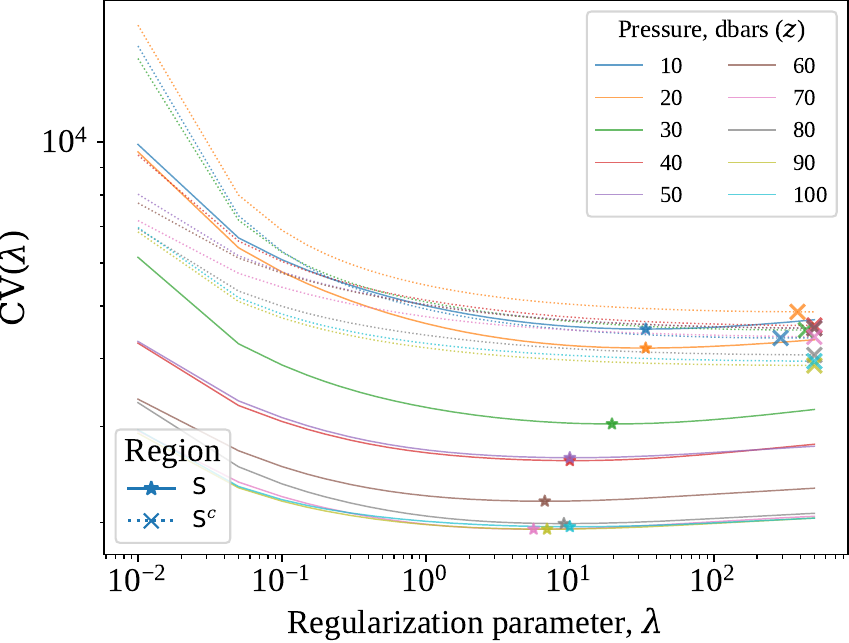}  
    \caption{
      $z \in \{10, 20, \dotsc, 100\}$
    }
		\label{fig:loocv-decomposed-1}
	\end{subfigure}
  \hspace{0.3cm}
	\begin{subfigure}{.48\textwidth}
		\centering
		\includegraphics[width=1.\linewidth]{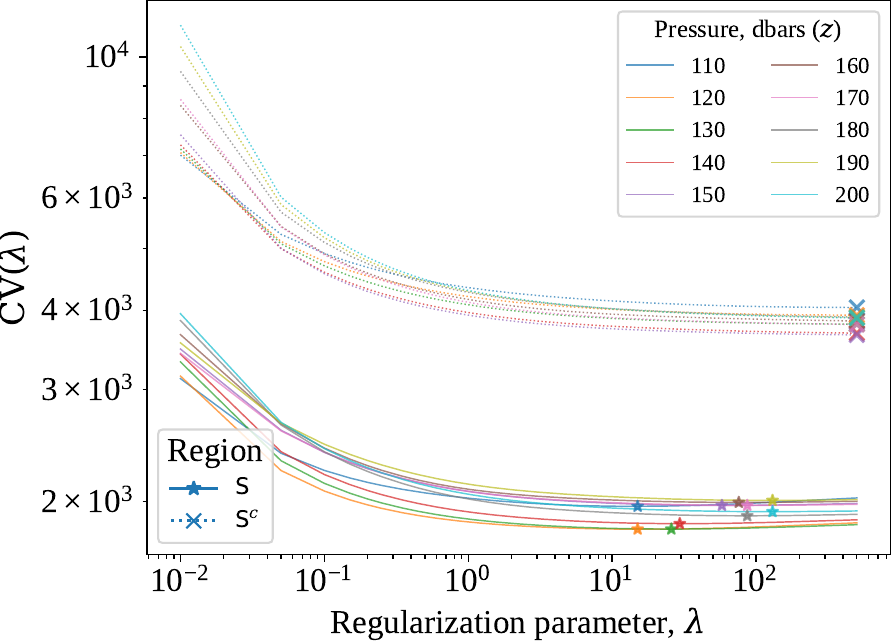}  
    \caption{
      $z \in \{110, 120, \dotsc, 200\}$
    }
    \label{fig:loocv-decomposed-2}
	\end{subfigure}
    \caption{
      LOOCV curves, computed separately using errors within a ``thermal
      response signal'' region $\mathcal{S}$ and errors within
      $\mathcal{S}^c$.  The observations within $\mathcal{S}$ consistently
      favor a more complex model; the fact that no single $\lambda$ is
      uniformly optimal over the entire domain highlights the heterogeneity
      of the signal being estimated.  We compromise by allowing a more
      complex model that achieves LOOCV error within 1\% of the minimum.
    }
	\label{fig:loocv-decomposed}
\end{figure}

Specifically, the signal is particularly large in magnitude for
cross-track angles ($\XTrack$) close zero and within the first two weeks after
the TC
passage ($\DT$); however, outside of this region, the signal drops
off and is close to zero in curvature.  For illustrative purposes, we
formally define this ``thermal response signal'' region to be
$(\XTrack, \DT) \in [-3, +3]\times[0, 12] =: \mathcal{S}$.  We then compute
two LOOCV
error curves for each depth $z$, one using data from $\mathcal{S}$ and the
other using data from $\mathcal{S}^c = \Omega\setminus\mathcal{S}$.

Indeed, we see in Figure~\ref{fig:loocv-decomposed} that the LOOCV curves
implied by the errors corresponding to observations in each of these two
regions produce very different minimizers, with the observations within the
``thermal response signal'' region favoring a more complex model, and
observations outside the signal region favoring a more regularized model.  The
choice of $\CVLamb = \hat\lambda^\text{1.01}$ represents a compromise between
the natural impulse to minimize the LOOCV error and the recognition that naive
minimization will yield an over-regularized model, as was empirically
confirmed to be the case when visually inspecting the fits.  
Importantly, the $\hat\lambda^{\mathcal{S}}$, tuned using only the
observations falling in $\mathcal{S}$, are roughly consistent with our slightly
undersmoothed $\CVLamb$, as can be seen comparing
Figures~\ref{fig:loocv-minimized-inflated} and
\ref{fig:loocv-decomposed}.

\noappendix       







\authorcontribution{
  AJH, MK, and ABL developed the methodology with domain knowledge and
  conceptualization from DG and KMW.  AJH performed the formal analysis,
  visualization, and writing, with feedback from all authors.  Software was
  developed by AJH extending original software by MK.
} 

\competinginterests{
	The authors declare that they have no conflict of interest.
} 


\begin{acknowledgements}
AJH acknowledges support from the NSF GRFP (Award DGE175016) and NSF DMS (Award
1520786). MK was supported by NOAA (Award NA21OAR4310258). DG was supported by
NSF EarthCube (Award 1928305) and NOAA (Award NA21OAR4310261). The authors
thank members of the CMU Statistical Methods for the Physical Sciences (STAMPS)
Research Group and the SAMSI/CMU Statistical Oceanography Working Group for
insightful comments and feedback during this work.

The Argo data used in this paper were collected and are made freely available
by the International Argo Program and the national programs that contribute to
it (\url{http://doi.org/10.17882/42182}).  
The Argo Program is part of the Global Ocean Observing System.
\end{acknowledgements}

\bibliographystyle{copernicus}
\bibliography{paper}

\end{document}